\documentclass[nojss]{jss}
\usepackage{thumbpdf,lmodern} 
\graphicspath{{Figures/}}

\usepackage{amsmath,subcaption}

\author{Marjolein Fokkema\\Universiteit Leiden}
\title{Fitting Prediction Rule Ensembles with \proglang{R} Package \pkg{pre}}

\Plainauthor{Marjolein Fokkema} 
\Plaintitle{pre: An R Package for Fitting Prediction Rule Ensembles} 
\Abstract{ Prediction rule ensembles (PREs) are sparse collections of
  rules, offering highly interpretable regression and classification
  models. This paper shows how they can be fitted using function
  \code{pre} from \proglang{R} package \pkg{pre}, which derives PREs
  largely through the methodology of \cite{FrieyPope08}. The
  implementation and functionality of \code{pre} is described and
  illustrated through application on a dataset on the prediction of
  depression. Furthermore, accuracy and sparsity of \code{pre} is
  compared with that of single trees, random forests, lasso regression
  and the original RuleFit implementation of \cite{FrieyPope08} in
  four benchmark datasets. Results indicate that \code{pre} derives
  ensembles with predictive accuracy similar to that of random
  forests, while using a smaller number of variables for
  prediction. Furthermore, \code{pre} provided better accuracy and
  sparsity than the original RuleFit implementation.  }

\Plainkeywords{prediction rules, ensemble learning, decision trees, lasso penalty, R}

\Address{
  Marjolein Fokkema\\
  Department of Methods and Statistics\\
  Institute of Psychology\\
  Faculty of Social Sciences\\
  Universiteit Leiden\\
  Leiden, The Netherlands\\
  E-mail: \email{m.fokkema@fsw.leidenuniv.nl}\
}

\begin{document}

\section{Introduction}
\label{sec:Introduction}

Prediction rule ensembles provide accurate and interpretable methods for regression and classification. Prediction rules are logical statements of the form \textit{if [conditions] then [prediction]}, which are easy to use in decision making. The prediction rules can be depicted as very simple decision trees, further improving interpretability \citep[e.g.,][]{FokkySmit15a}. 

Several algorithms for deriving decision trees are available, an early
example being the classification and regression tree algorithm
\citep[CART;][]{BreiyFrie84}. Although CART trees are easy to
interpret, they suffer from two disadvantages: biased variable
selection and instability. Although the variable selection bias has
been addressed by several later tree induction algorithms, like for
example the conditional inference trees algorithm of
\cite{HothyHorn06}, the problem of instability is shared by all tree
induction algorithms. Instability here means that small changes in the
training data may yield major changes in the resulting tree.

A powerful solution to the instability problem is combining the
predictions of single trees through ensembling, which has been found
to substantially improve predictive accuracy \citep[e.g.,][]{Brei96,
  Diet00, StroyMall09}. However, the resulting ensembles generally
consist of a large number of trees and are therefore difficult to
interpret and apply. A trade-off between accuracy and interpretability
seems to apply: single trees provide better interpretability, whereas
tree ensembles provide better accuracy.

Prediction rule ensembles (PREs) aim to optimize accuracy as well as
interpretability, by creating ensembles with a small number of simple
trees or rules. Several algorithms for deriving PREs have been
developed, most exclusively aimed at classification, like SLIPPER
\citep{CoheySing99} and lightweight rule induction
\citep{WeisyIndu00}. Alternatively, the RuleFit \citep{FrieyPope08},
ENDER \citep{DembyKotl10} and node harvest \citep{Mein10} algorithms
can be applied to classification as well as regression problems. The
RuleFit algorithm generates a large initial ensemble of rules from a
boosted tree ensemble and selects a sparse final rule ensemble using
lasso regression. This approach yields ensembles that are competitive
in accuracy with, for example, boosted tree ensembles and random
forests \citep{FrieyPope08, JolyySchn12, ShimyLi14, YangyZhan08}.

The aim of the current paper is to introduce function \code{pre} from
\proglang{R} \citep{R19} package \pkg{pre} \citep{FokkyChri19}, which
provides a completely
 \proglang{R}-based implementation of the
algorithm of \cite{FrieyPope08}. Package \pkg{pre} is available from
the Comprehensive \proglang{R} Archive Network (CRAN) at
\url{https://CRAN.R-project.org/package=pre}.  Although the
\pkg{RuleFit} program \citep{FrieyPope12} already provides a fast
implementation of the algorithm, \code{pre} provides a number of
potential advantages: First, in addition to the CART algorithm, it
allows for employing the unbiased recursive partitioning algorithms of
\cite{HothyHorn06} and \cite{ZeilyHoth08} to generate rules. Second,
in addition to continuous and binary outcomes, \code{pre} allows for
the analysis of count, multinomial, multivariate continuous and
survival outcomes. Third, in addition to bagging and boosting,
\code{pre} allows for generating prediction rules through a
random-forest style approach. Fourth, \code{pre} is completely
\proglang{R}-based, providing \proglang{R} users with a more familiar
interface and more easily accessible results and documentation. The
first and last advantages, however, come at a computational cost,
yielding longer computation times for \code{pre} than for the original
\pkg{RuleFit} program.

In what follows, the implementation (Section~\ref{sec:Implementation})
and functionality (Section~\ref{sec:Usage}) of \code{pre} will be
described. In Section~\ref{sec:Example}, application of \code{pre}
will be illustrated using an existing dataset on the prediction of
depressive symptomatology. Also, several examples will illustrate how
well-known tree ensemble approaches can be mimicked in rule
generation. In Section~\ref{sec:Empirical}, the performance of
\code{pre} will be compared with that of single trees, random forests,
lasso penalized linear regression and the original RuleFit
implementation in four benchmark datasets. In
Section~\ref{sec:Comparison}, the properties of \code{pre} and other
software packages for deriving PREs will be discussed and compared.

\section{Implementation}
\label{sec:Implementation}

\subsection{Rule generation}

Following the methodology of \cite{FrieyPope08}, \code{pre} first
generates a large ensemble of decision trees, from which an initial
ensemble of prediction rules is derived. To induce trees, \code{pre}
by default employs function \code{ctree} from the \pkg{partykit}
package \citep{HothyZeil15}. Alternatively, function \code{glmtree}
\citep[also from package \pkg{partykit};][]{ZeilyHoth08} or function
\code{rpart} from package \pkg{rpart} \citep{TheryAtki19} can be
employed. Function \code{rpart} implements the original CART algorithm
of \cite{BreiyFrie84}, whereas functions \code{ctree} and
\code{glmtree} implement unbiased recursive partitioning procedures,
which address the variable selection bias mentioned earlier, through
separating variable and cut-point selection. Function \code{glmtree}
fits GLMs with different parameter estimates in every node. To obtain
a tree with constant fits in the nodes with \code{glmtree}, an
intercept-only GLM is specified.

Functions \code{ctree}, \code{glmtree} and \code{rpart} employ
different criteria for variable and/or cut-point selection and will
yield somewhat different tree structures, given the same
data. Although \code{glmtree} will yield the longest computation
times, it may yield slightly higher accuracy. For further details on
variable and cut-point selection criteria employed by the different
algorithms, the reader is referred to \cite{HothyHorn06},
\cite{ZeilyHoth08} and \cite{TheryAtki19}.

\begin{figure}[t!]
\centering
\includegraphics[width=0.6\textwidth, trim = 0 25 0 20, clip]{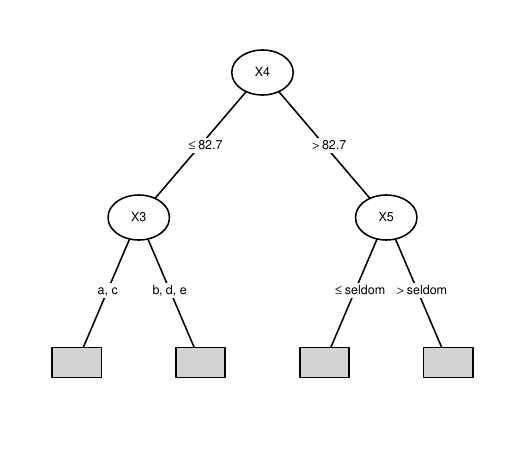}
\caption{Example tree. Information on the distribution of the outcome
  variable in the terminal nodes is omitted, as only the tree
  structure is used for generating rules.}
\label{fig:tree}
\end{figure}

To illustrate rule generation, Figure~\ref{fig:tree} depicts an
example tree. From this tree, the following set of rules can be
derived:
\begin{align*}
  r_1(\mathbf{x}) &= I (x_4 \leq 82.7),\\
  r_2(\mathbf{x}) &= I ((x_4 \leq 82.7) \cdot (x_3 \in \{a,c\})),\\
  r_3(\mathbf{x}) &= I ((x_4 \leq 82.7 \cdot (x_3 \in \{b,d,e\})),\\
  r_4(\mathbf{x}) &= I (x_4 > 82.7),\\
  r_5(\mathbf{x}) &= I ((x_4 > 82.7) \cdot (x_5 \leq \textrm{seldom})),\\
  r_6(\mathbf{x}) &= I ((x_4 > 82.7) \cdot (x_5 > \textrm{seldom})),
\end{align*}
where $r_k(\mathbf{x})$ denotes prediction rule $k$
($k = 1, \dots, K$), taking a value of 1 if its conditions apply, and
a value of 0 if not; $\mathbf{x}$ denotes a random vector of $p$ input
variables; $x_j$ denotes input variable $j$ ($j = 1,\dots, p$); and
$I$ is a function denoting the truth of its argument. Note that, with
exception of the root node, a rule is generated for every node in the
tree, not just for the terminal nodes. This also yields redundant
rules: for example, the first and fourth rule above are perfectly
collinear, that is $r_4(\mathbf{x}) = 1 - r_1(\mathbf{x})$. Therefore
$r_4$ will be omitted from the initial ensemble. Similarly, rules that
are identical to earlier generated rules are also removed from the
initial ensemble. Furthermore, as shown in Figure~\ref{fig:tree},
input variables may be continuous (like $x_4$), unordered categorical
(like $x_3$) or ordered categorical (like $x_5$).

The ensemble of decision trees can be generated in an approach similar
to bagged, boosted or random forest tree ensembles, or a combination
thereof. By default, \code{pre} draws $B = 500$ random sub-samples
from the training data and grows a tree on each sample. As in bagging,
\code{pre} allows for employing bootstrap sampling (i.e., sampling
with replacement), but sub-sampling has been found to yield lower
inclusion frequencies for noise variables \citep{DeBiyJani16}. In
addition to random sampling of observations, \code{pre} allows for
random-forest style sampling of predictor variables for split
selection through specification of an \code{mtry} argument. By
default, however, all predictor variables are considered for split
selection.

To apply boosting, a learning rate (or shrinkage parameter) $\nu$ can be specified, which controls the influence of earlier trees on the induction of new trees. If $\nu > 0$, a gradient boosting approach is employed, where the tree in iteration $b$ ($b = 1, \dots, B$) is grown on the pseudo response $\tilde{y}_b$, instead of the original response $y$. 

For a continuous response variable $y$, the pseudo response in iteration $b$ is given by:
\begin{equation}
  \tilde{y}_b  = y - \sum_{m=1}^{b-1}{\nu \cdot f_m(\mathbf{x})},
  \label{eq:pseudo_y_continuous}
\end{equation}
where $f_m(\mathbf{x})$ are the predictions from the regression tree
fitted in iteration $m$ ($m = 1, \dots, b-1$). For a multivariate
continuous response variable $\mathbf{y}$,
Equation~\ref{eq:pseudo_y_continuous} would also be employed, but with
a multivariate pseudo response $\tilde{\mathbf{y}}_b$ and multivariate
predictions $f_m(\mathbf{x})$. For a binary (0$-$1 coded) response
variable, regression trees are fitted to a continuous pseudo response,
which is given by:
\begin{equation}
  \tilde{y}_b = \frac{y - p_{b-1}}{\sqrt{p_{b-1}(1 - p_{b-1})}} \; \mathrm{,}  
  \label{eq:pseudo_y_binary}
\end{equation}
where 
\begin{equation}
  p_{b-1} = \frac{1}{1 + e^{-\eta_{b-1}}} \mathrm{.} 
\end{equation}
In the first iteration, the value of $\eta$ is given by:
\begin{equation}
  \eta_{0} = \mathrm{log} \left( \frac{\bar{p}}{1-\bar{p}} \right) \mathrm{,}
\end{equation}
where $\bar{p}$ is the (possibly weighted) mean of $y$. In subsequent iterations ($b > 1$), $\eta$ is given by:
\begin{equation}
  \eta_{b-1} = \eta_{0} + \sum_{m=1}^{b-1}{\nu \cdot f_m(\mathbf{x})} \mathrm{.}  
  \label{eq:eta_b_minus_one}
\end{equation}
For multinomial response variables, gradient boosting is applied by
coding the response as a set of 0$-$1 coded indicator variables, one for
each response category, and applying
Equations~{\ref{eq:pseudo_y_binary}}
through {\ref{eq:eta_b_minus_one}} to
obtain a multivariate pseudo response.

For count response variables, regression trees are also employed and the pseudo response is given by:
\begin{equation}
  \tilde{y}_b = y - \lambda_{b-1} \mathrm{,}  
\end{equation}
where 
\begin{equation}
  \lambda_{b-1} = e^{\eta_{b-1}}  \mathrm{.} 
\end{equation}
In the first iteration, the value of $\eta$ is given by:
\begin{equation}
  \eta_{0} = \mathrm{log} \left( \bar{\lambda}  \right) \mathrm{,} 
\end{equation}
where $\bar{\lambda}$ is the (possibly weighted) mean of $y$. In
subsequent iterations $(b>1)$, $\eta$ is given by
Equation~{\ref{eq:eta_b_minus_one}}.

For the gradient boosting approach described above, functions
\code{ctree} or \code{rpart} are employed. Alternatively, function
\code{glmtree} can be employed, allowing for application of the
learning rate through including an offset in the GLM (i.e., a
predictor with a fixed coefficient of 1). That is, in every iteration,
a GLM-based recursive partition is fit on the response $y$ with the
offset in each iteration $b$ given by:
\begin{equation}
  \eta_{b-1}  = \sum_{m=1}^{b-1}{\nu \cdot f_m(\mathbf{x})} \mathrm{,} 
\end{equation}
where $f_m(\mathbf{x})$ are the predictions on the scale of the linear
predictor from the tree fitted in iteration $m$. Whereas gradient
boosting with \code{ctree} or \code{rpart} yields shorter computation
times, the use of \code{glmtree} may yield a somewhat more accurate
final ensemble.

Based on the findings of \cite{Frie01} and \cite{FrieyPope03}, who
found small non-zero values of the learning rate to perform best for
ensembles of small decision trees, the learning rate $\nu$ of
\code{pre} is set to 0.01, by default.

Although the original bagging and random forest algorithms made use of
unpruned trees, limiting tree size has been found to yield better
predictive accuracy \citep[e.g.,][]{LinyJeon06}. Smaller trees also
yield shorter prediction rules, which are easier to
interpret. Function \code{pre} therefore generates trees with a
maximum depth of three by default, yielding a maximum of three
conditions per rule and restricting interactions that can be captured
to first- and second-order ones. Other values for maximum tree depth
can be specified by the user.

\subsection{Selection of the final ensemble}

After the ensemble of decision trees is generated, every node from every tree is included as a rule in the initial rule ensemble. Rules that are perfectly collinear with earlier rules are omitted from the initial ensemble, by default. Furthermore, all predictor variables are included as linear terms in the initial ensemble, by default. This may improve sparsity and/or accuracy of the final ensemble, as rules may have difficulty approximating purely linear functions of input variables. Alternatively, the ensemble may be specified to include either rules or linear terms only. To reduce the effect of possible outliers, continuous and ordered categorical predictor variables are winsorized before inclusion as linear terms in the initial ensemble: 
\begin{equation}
  l_j(x_j) = \min(\delta_j^+, \max(\delta_j^-, x_j)),
  \label{eq:linear_terms}	
\end{equation}
where $\delta_j^-$ and $\delta_j^+$ represent the $\beta$ and
$(1-\beta)$ quantiles of the distribution of predictor variable $x_j$
in the training data. By default, $\beta$ is set to 0.025, but other
values may be specified by the user.

Unordered factors are included in the initial ensemble as $(q_{j}-1)$
0$-$1 coded variables, where $q_j$ corresponds to the number of levels
of predictor variable $j$. The resulting initial ensemble consists of
a large number of base learners (rules and/or linear terms), of which
only a small subset may actually contribute to predictive
accuracy. Therefore, coefficients for the base learners are estimated
using penalized regression. By default, \code{pre} employs the lasso
penalty, but ridge or elastic net penalties can also be selected by
the user.

As the lasso penalty more heavily penalizes predictors with smaller variance, linear terms are normalized before estimation, by default:
\begin{equation} 
  l_j(x_j) \leftarrow 0.4 \cdot \frac{l_j(x_j)}{\mathit{sd}(l_j(x_j))},
  \label{eq:normalizing}
\end{equation}
where $sd(l_j(x_j))$ is the sample standard deviation of the linear
term (Equation~{\ref{eq:linear_terms}}) and 0.4 represents the
standard deviation of a typical rule.\footnote{The standard deviation
  of a typical rule is derived as follows: A rule is a binary 0$-$1
  coded variable. If we take $s$ to be the proportion of observations
  to which a rule applies, the variance is given by $f(s) = s(1-s)$,
  with antiderivative
  $F(s) = \int s(1-s) \mathrm{d}s = 1/2 \cdot s^2 - 1/3 \cdot
  s^3$. The average of a function $f(s)$ over its domain $[a,b]$ is
  given by $1/(b-a) \int_{a}^{b} f(s) \mathrm{d}s$. Thus, the average
  variance of a rule is equal to
  $1/(1-0) \int_{0}^{1} f(s) \mathrm{d}s = \int_{0}^{1} f(s)
  \mathrm{d}s = F(1) - F(0) = (1/2 - 1/3) - (0 - 0) = 1/6$, yielding a
  standard deviation of $\sqrt{1/6} \approx 0.4$.}

If both rules and linear terms are included in the ensemble, the final predictive model is given by:
\begin{equation}
  F(\mathbf{x}) = \hat{a}_0 + \sum_{k=1}^K{\hat{a}_k r_k(\mathbf{x}) + \sum_{j=1}^p{\hat{b}_j l_j(x_j)}}.
  \label{eq:predictive_model}
\end{equation} 
If the lasso penalized regression is employed, coefficients $\hat{a}$ and $\hat{b}$ are estimated by minimizing:
\begin{equation}
  \sum^{N}_{i=1} {L \left( y_i, a_0 + \sum^{K}_{k=1}{a_k r_k(\textbf{x}_i)} + \sum^{p}_{j=1}{b_j l_j(x_{i,j})} \right)} + \lambda \cdot \left( \sum^{K}_{k=1}{|a_k| + \sum^p_{j=1}{|b_j|}} \right),
  \label{eq:lasso}
\end{equation}
where $i$ ($i = 1, \dots, N$) denotes an observation in the training
dataset. For estimation of coefficients, \code{pre} employs the
\code{cv.glmnet} function from package \pkg{glmnet}
\citep{FrieyHast10}. By default, the loss function $L$ is equal to 0.5
times the squared residual for continuous responses and minus the
log-likelihood for other response types. Other loss functions (e.g.,
mean absolute error, misclassification error) may be specified by the
user. By default, the penalty parameter $\lambda$ is set to the value
yielding a cross-validated prediction error of one standard error
within the minimum, but other values for $\lambda$ may also be
specified by the user.

\subsection{Interpretation}

\cite{FrieyPope08} proposed several measures for interpretation of the
final prediction rule ensemble. Most are implemented in \pkg{pre} and
discussed below.

\subsubsection{Importance}

To quantify the relative contribution of every base learner to the
predictions of the final ensemble, importances can be
calculated. \cite{FrieyPope08} defined the importance of a linear term
as:
\begin{equation}
  I_j = |\hat{b}_j| \cdot \mathit{sd}(l_j(x_j)), % equation 29 in F\&P
  \label{eq:global_linear_imp}
\end{equation}
where $\mathit{sd}$ denotes the sample standard deviation. Similarly, the global importance of a rule is given by:
\begin{equation}
  I_k = |\hat{a}_k| \cdot \sqrt{s_k(1-s_k)}, % equation 28 in F\&P
  \label{eq:global_rule_imp}
\end{equation}
where $\sqrt{s_k(1-s_k)}$ is the sample standard deviation of rule
$k$, which is in turn defined by $s_k$, the support of rule $k$ in the
training data, or the proportion of training observations to which
rule $k$ applies:
\begin{equation}
  s_k = \frac{1}{N} \sum^{N}_{i=1}{r_k(\mathbf{x}_i)}.
  \label{eq:global_support}
\end{equation}

The importances in Equations~\ref{eq:global_linear_imp} and
\ref{eq:global_rule_imp} can be interpreted as the absolute value of
regression coefficients, standardized with respect to the base
learner. Additional standardization with respect to the outcome
variable would yield importances that can be interpreted as
standardized regression coefficients. Therefore, \pkg{pre} also allows
for calculating standardized importances for continuous
outcomes. Thus, for linear terms, the standardized importance is given
by:
\begin{equation}
  I'_j = |\hat{b}_j| \cdot \frac{\mathit{sd}(l_j(x_j))}{\mathit{sd}(y)}.
  \label{eq:global_linear_imp_stand}
\end{equation}
For rules, the standardized importance is given by:
\begin{equation}
	I'_k = |\hat{a}_k| \cdot \frac{\sqrt{s_k(1-s_k)}}{\mathit{sd}(y)}.
	\label{eq:global_rule_imp_stand}
\end{equation}
The total importance of input variable $x_j$ is given by the sum of the importances of the linear term and the rules in which $x_j$ appears \citep{FrieyPope08}:
\begin{equation}
  J_j = I_j + \sum_{x_j \in r_k}\frac{I_k}{c_k},
  \label{eq:glob_inputvar_imp}
\end{equation}
where $c_k$ is the number of conditions that define rule $k$. The
second term $\sum_{x_j \in r_k}{I_k / c_k}$ shows that the importance
of a rule is distributed equally over the input variables appearing in
the rule. When a variable appears more than once in the conditions of
a rule, $I_k / c_k$ is multiplied accordingly. The variable importance
in Equation~{\ref{eq:glob_inputvar_imp}} can be calculated using
standardized or unstandardized importances. In either case, variable
importances take values $\geq 0$, with higher values indicating a
stronger effect on the ensemble's predictions.

\cite{FrieyPope08} also proposed local importance measures for a
selected subregion of the input variable space. These can be obtained
by replacing the global standard deviations in
Equations~{\ref{eq:global_linear_imp}} and {\ref{eq:global_rule_imp}}
(or {\ref{eq:global_linear_imp_stand}} and
{\ref{eq:global_rule_imp_stand}} for their standardized counterparts)
by the local standard deviations in the subregion. Local variable
importances can in turn be calculated by summing the weighted local
importances of the base learners in which the variable appears, as in
Equation~{\ref{eq:glob_inputvar_imp}}.

\subsubsection{Partial dependence}

The shape of the association between predictor and response variables
can be assessed through plotting partial dependence functions. The
partial dependence of $F(\mathbf{x})$ on a subset of input variables
$S \subset \{1, \ldots, p\}$ is defined as the expected value of
$F(\mathbf{x})$ over the marginal joint distribution of input
variables not in $S$ (i.e., $\mathbf{x}_{\setminus S}$;
\citealt{Frie01, FrieyPope08}). The partial dependence of
$F(\mathbf{x})$ on the subset of predictor variables $S$ can be
estimated from the data by:
\begin{equation}
  \hat{F}_S(\mathbf{x}_S) = \frac{1}{N} \sum^N_{i=1}{F(\mathbf{x}_S, \mathbf{x}_{i, \setminus S})}, % equation 40 in F/&P
  \label{eq:part_dep}
\end{equation}
where $\{ \mathbf{x}_{i, \setminus S} \}^N_{i=1}$ are the training
data observations. Taking a subset $S$ of one or two predictor
variables allows for plotting the partial dependence of
$F(\mathbf{x})$ on $\mathbf{x}_S$.

\subsubsection{Interactions}

Prediction rules are well suited for capturing interaction effects of input variables. However, non-zero coefficients for rules involving multiple predictor variables in the final ensemble are a necessary but not sufficient condition for the presence of interaction effects. For example, the interaction may be cancelled out by other rules in the ensemble. Or, a rule involving multiple predictor variables may merely reflect main effects of (correlated) predictor variables, instead of interaction(s).

\cite{FrieyPope08} developed a statistic for assessing whether a
predictor variable is involved in interactions with other predictor
variables in the model. The underlying rationale is that in the
presence of interaction effects, the effects of individual predictor
variables are not additive. If an input variable $x_j$ is not involved
in interactions with other input variables $\mathbf{x}_{\setminus j}$,
then its effect on $F(\mathbf{x})$ is additive, which can then be
expressed as:
\begin{equation}
  F(\mathbf{x}) = F_j(x_j) + F_{\setminus j}(\mathbf{x}_{\setminus j}), % equation 42
  \label{eq:additive_effects}
\end{equation}
where $F_j(x_j)$ is the partial dependence of $F(\mathbf{x})$ on $x_j$
and $F_{\setminus j}(\mathbf{x}_{\setminus j})$ is the partial
dependence of $F(\mathbf{x})$ on $\mathbf{x}_{\setminus j}$, both of
which can be estimated from the data using
Equation~{\ref{eq:part_dep}}. If we assume all
partial dependence functions as well as the predictive model
$F(\mathbf{x})$ to be centered to have a mean value of zero, the
extent to which $F(\mathbf{x})$ deviates from additivity with respect
to $x_j$ can be quantified though the following statistic:
\begin{equation}
	H^2_j = \frac{\sum^N_{i=1}{ [F(\mathbf{x}_i) - \hat{F}_j(x_{i,j}) - \hat{F}_{\setminus j}(\mathbf{x}_{i, \setminus j})]^2 }}{\sum^N_{i=1}{ [F(\mathbf{x}_i)]^2}},
	\label{eq:interaction_statistic}
\end{equation}
where $F(\mathbf{x}_i)$ represent the (centered) model predictions at
$\mathbf{x}_i$; $\hat{F}_j(x_{i,j})$ represents the (centered) partial
dependence of the predictive model on $x_j$, evaluated at $x_{i,j}$;
$\hat{F}_{\setminus j} (\mathbf{x}_{i, \setminus j})$ represents the
(centered) partial dependence of the predictive model on
$\mathbf{x}_{\setminus j}$, evaluated at $x_{i, \setminus j}$ and the
denominator $\sum^N_{i=1}{ [F(\mathbf{x}_i)]^2}$ represents the sample
variance of the model's predictions. The statistic $H^2_j$ measures
the fraction of variance of $F(\mathbf{x})$ not captured by the
additive effects. It will differ from zero to the extent that $x_j$ is
involved in interactions with other input variables.

To assess whether the estimated $H^2_j$ value significantly differs
from zero, a null distribution has to be derived. \cite{FrieyPope08}
suggest the use of a variant of the parametric bootstrap
\citep{EfroyTibs94} to derive a null distribution for $H_j^2$. In
effect, $H^2_j$ is repeatedly computed for ensembles fitted on
artificial datasets from which interactions are known to be
absent. The procedure for generating artificial datasets without
interactions and calculating the reference distribution for $H_j^2$ is
described in detail in \citet[][Section 8.3]{FrieyPope08}.

\section{Usage}
\label{sec:Usage}

The basic usage and default settings of function \code{pre} are as follows:
\begin{Code}
pre(formula, data, weights, family = gaussian, use.grad = TRUE, 
  tree.unbiased = TRUE, type = "both", sampfrac = 0.5, maxdepth = 3L, 
  learnrate = .01, confirmatory = NULL, mtry = Inf, ntrees = 500, 
  tree.control, removeduplicates = TRUE, removecomplements = TRUE, 
  winsfrac = 0.025, normalize = TRUE, standardize = FALSE,
  ordinal = TRUE, nfolds = 10L, verbose = FALSE, par.init = FALSE, 
  par.final = FALSE, ...)
\end{Code}
The following arguments are required:

\begin{itemize}
 \setlength\itemsep{0pt}
\item \code{formula} provides a symbolic description of the model to
  be fit.
\item \code{data} specifies a data frame containing the variables in
  the model.
\end{itemize}

The following arguments are optional:

\begin{itemize}
\setlength\itemsep{0pt}

 \item \code{weights} provides a vector of case weights.
 
 \item \code{family} specifies a GLM-family object or a character
   string. By default, \code{family = gaussian} and a single
   continuous response variable is assumed. Alternatively, for a
   single binary factor \code{binomial}, for a count response
   \code{poisson}, for a factor with $>2$ levels \code{"multinomial"},
   for a multivariate continuous response \code{"mgaussian"} and for a
   survival response \code{"cox"} should be specified. Note that
   gaussian, binomial and poisson families may be specified as either
   a GLM-family object or a character string.
 
 \item \code{use.grad} specifies whether a gradient boosting approach
   should be employed to apply the learning rate. If set to
   \code{FALSE}, \code{glmtree} instead of \code{ctree} or
   \code{rpart} is employed for tree induction.
 
 \item \code{tree.unbiased} specifies whether an unbiased tree generation algorithm should be employed for rule generation. \code{TRUE} by default, if set to \code{FALSE}, function \code{rpart} will be employed for tree induction.

 \item \code{type} specifies the type of base learners to be included in the ensemble: \code{"rules"}, \code{"linear"} or \code{"both"}. 
 
 \item \code{sampfrac} specifies the fraction of training observations
   sampled to produce each tree. Values $<1$ yield sampling with
   replacement (sub-sampling), a value of 1 yields sampling with
   replacement (bootstrap sampling). Alternatively, a sampling
   function may be supplied, which should take arguments \code{n} and
   \code{weights}.
 
 \item \code{maxdepth} specifies the maximum tree depth and thereby the maximum number of conditions in rules. Should be an integer of length 1 or \code{ntrees}. Alternatively, a random number generating function may be supplied, which should take argument \code{ntrees}. 
 
 \item \code{learnrate} specifies the value of the learning rate $\nu$ to be applied in tree induction.

 \item \code{confirmatory} specifies a character vector of
   confirmatory terms to be included in the final ensemble. No penalty
   will be applied to the coefficients of these terms, which will
   yield a non-zero coefficient for the term in the final ensemble.
   
 \item \code{mtry} specifies the number of randomly selected predictor
   variables for selecting splits in trees. The default \code{Inf}
   yields no prior selection of predictor variables.
 
 \item \code{ntrees} specifies the number of trees to be grown for deriving the initial ensemble of trees.
 
 \item \code{tree.control} specifies a list of additional control parameters to be passed to the tree induction algorithm.
 
 \item \code{removeduplicates} specifies whether rules which are identical to earlier generated rules (that is, apply to the same set of observations) should be removed from the initial ensemble.
 
 \item \code{removecomplements} specifies whether rules which are the exact complement of earlier rules (that is, are equal to 1 minus an earlier rule) should be removed from the initial ensemble. 
 
 \item \code{winsfrac} specifies the quantiles of the data distribution to be used for winsorizing linear terms. If set to 0, no winsorizing is performed.
 
 \item \code{normalize} specifies whether linear terms should be
   normalized before estimation of the final ensemble. The default
   results in every linear term being scaled to have a standard
   deviation of 0.4, equal to that of a typical rule.
 
 \item \code{standardize} specifies whether all rules and linear terms should be standardized to have unit variance before estimation of the final ensemble.
 
 \item \code{nfolds} specifies the number of folds to be used in calculating the cross-validated error estimates for the possible penalty parameter value for selection of the final ensemble.
 
 \item \code{ordinal} specifies whether ordered factors should be
   treated as continuous variables for generating rules. The default
   generally yields simpler rules and computation time.
 
 \item \code{verbose} specifies whether information on model fitting progress should be printed to the command line.
 
 \item \code{par.init} specifies whether parallel computation should be employed for fitting the initial tree ensemble. Note that parallel computation of the initial ensemble will reduce computation time only for (very) large datasets.
 
 \item \code{par.final} specifies whether parallel computation should be employed for selecting the final ensemble.
 
 \item \code{...} specifies additional arguments to be passed to function \code{cv.glmnet}.
 
\end{itemize}

The generated ensemble is returned as an object of class `\code{pre}',
which offers several standard methods and functions for extracting
information.

Note that the default settings of \code{pre} represent the author's
settings of choice, which tend to favor relatively sparse
ensembles. The many arguments of function \code{pre} allow users to
carefully tune accuracy and sparsity of the final ensemble, or to
mimic existing tree ensemble approaches for generating rules (e.g.,
bagging, random forests). Several examples illustrating the tuning
options will be provided in
Section~{\ref{subsec:custom}}. First, the
next section will illustrate the functionality of \code{pre} in a
real-data example with default settings.

\section{Examples}
\label{sec:Example}

\subsection{Prediction of depression}

To illustrate application of \code{pre}, we use a dataset from a study
by \cite{CarryRojo01} which is included in the package. This study
examined the extent to which subscales of the NEO personality
inventory \citep[NEO-PI;][]{CostyMcCr85} are predictive of depressive
symptomatology as measured by the Beck depression inventory
\citep[BDI;][]{BeckyStee88}. The NEO-PI assesses five major
personality dimensions: neuroticism, extraversion, openness to
experience, agreeableness and conscientiousness. Each of these
dimensions is quantified through six subscale scores and one total
score. In the study of \cite{CarryRojo01}, the NEO-PI and BDI were
administered to 112 Spanish respondents. Total scores were calculated
for each of the six major dimensions, as well as for each of the
neuroticism, extraversion and openness subscales. Respondents' age in
years and sex were also included in the dataset. Further information
about the study and sample is provided in \cite{CarryRojo01}.

First, we load the package and data:
\begin{Schunk}
\begin{Sinput}
R> library("pre")
R> data("carrillo", package = "pre")
\end{Sinput}
\end{Schunk}
We derive a prediction rule ensemble using function \code{pre}. As rule derivation and selection of the final ensemble depends on random sampling of the training data, we first set the random seed:
\begin{Schunk}
\begin{Sinput}
R> set.seed(42)
R> carrillo.ens <- pre(bdi ~ ., data = carrillo)
R> carrillo.ens
\end{Sinput}
\begin{Soutput}
Final ensemble with cv error within 1se of minimum: 
  lambda =  0.7779287
  number of terms = 14
  mean cv error (se) = 37.05145 (6.212951)

  cv error type : Mean-Squared Error

         rule  coefficient               description
  (Intercept)   8.93985818                         1
       rule80   2.40103417     n4 > 15 & open4 <= 13
       rule88  -1.35678309     ntot <= 109 & e6 > 15
       rule97  -1.20824458     n2 <= 16 & open4 > 10
       rule18  -1.02385159  ntot <= 109 & etot > 101
        rule1  -0.99590845                  n3 <= 17
       rule12  -0.58728116                  n3 <= 22
       rule42  -0.51972664     n6 <= 19 & open4 > 12
       rule86  -0.26280878        n2 <= 16 & e6 > 14
       rule66   0.25137293   open4 <= 13 & ntot > 82
           n3   0.17522150         2 <= n3 <= 30.225
      rule105  -0.14432622     n2 <= 16 & open5 > 11
       rule30  -0.06473571    ntot <= 109 & n4 <= 14
       rule46  -0.05753864                  n1 <= 20
       rule40  -0.04878180                  n6 <= 19
\end{Soutput}
\end{Schunk}
The printed result reports that the final ensemble with
cross-validated error within one standard error above the minimum was
selected. This is the default employed by \code{print} and other
functions in \pkg{pre}. Alternatively, the
\code{penalty.parameter.val} argument can be set to
\code{"lambda.min"} ($\lambda$ value yielding the minimum
cross-validated error), or a numeric value $> 0$. Note that the
reported cross-validated error is calculated using the same data as
used for deriving the prediction rules and likely provides an overly
optimistic estimate of future prediction error. Performing full
cross-validation will yield a more honest estimate of prediction
error, for which we will use function \code{cvpre} later on in the
example.

The printed result shows each of the selected base learners in the
final ensemble with the corresponding coefficients. Base learners with
an estimated coefficient of 0 are omitted from this output, by
default. The first column (\code{rule}) indicates the type of base
learner: a rule (e.g., \code{rule80}) or linear term (e.g.,
\code{n3}). The \code{description} column lists the conditions for
rules and the winsorizing points for linear terms, if winsorizing was
performed (note that \code{n3} was winsorized with the default value
of $\beta = 0.025$). The first rule shows that observations with a
higher value of \code{n4} (i.e., n4 $>15$) and a lower value of
\code{open4} (i.e., open4 $\leq 13$) have an expected BDI score 2.4
higher than observations that do not match these conditions.

The results indicate that depressive symptomatology is mostly
predicted by the neuroticism (sub)scales. This is not surprising,
given that these scales were specifically constructed to assess
emotional adjustment and (in)stability, with higher scores indicating
higher proneness to psychological distress. The \code{ntot} variable
reflects the total score on this scale, while the \code{n1},
\code{n2}, \code{n3}, \code{n4} and \code{n6} variables reflect the
scores on the anxiety, anger \& hostility, depression,
self-consciousness and vulnerability subscales,
respectively. Furthermore, variable \code{e6} represents a subscale of
the extraversion scale, reflecting proneness to the experience of
positive emotions. The \code{open4} and \code{open5} variables
represent subscales of the openness to experience scale, capturing
openness to actions and ideas, respectively.

We can obtain the estimated (zero and non-zero) coefficients for the base learners in the final ensemble using the \code{coef} method (results not shown for space considerations):
\begin{Schunk}
\begin{Sinput}
R> coef(carrillo.ens)
\end{Sinput}
\end{Schunk}
We can obtain predictions for (new) observations using the \code{predict} method (results not shown for space considerations):
\begin{Schunk}
\begin{Sinput}
R> predict(carrillo.ens, newdata = carrillo)
\end{Sinput}
\end{Schunk}
We can obtain variable and base learner importances using the \code{importance} function. By default, importances are calculated over all training observations, but the \code{importance} function also allows for obtaining local importances calculated over a subset of the training data, through specification of the \code{global} and \code{quantprobs} arguments. To aid in interpretation, we request standardized importances in this example, so we can interpret the base learner importances as the absolute value of standardized multiple regression coefficients. Also, we restrict the number of decimal places in the results by specifying the \code{round} argument: 
\begin{Schunk}
\begin{Sinput}
R> imps <- importance(carrillo.ens, standardize = TRUE, round = 4L)
\end{Sinput}
\end{Schunk}
\begin{figure}[t!]
\centering
\includegraphics[width=0.7\textwidth, trim = 0 12 0 42, clip]{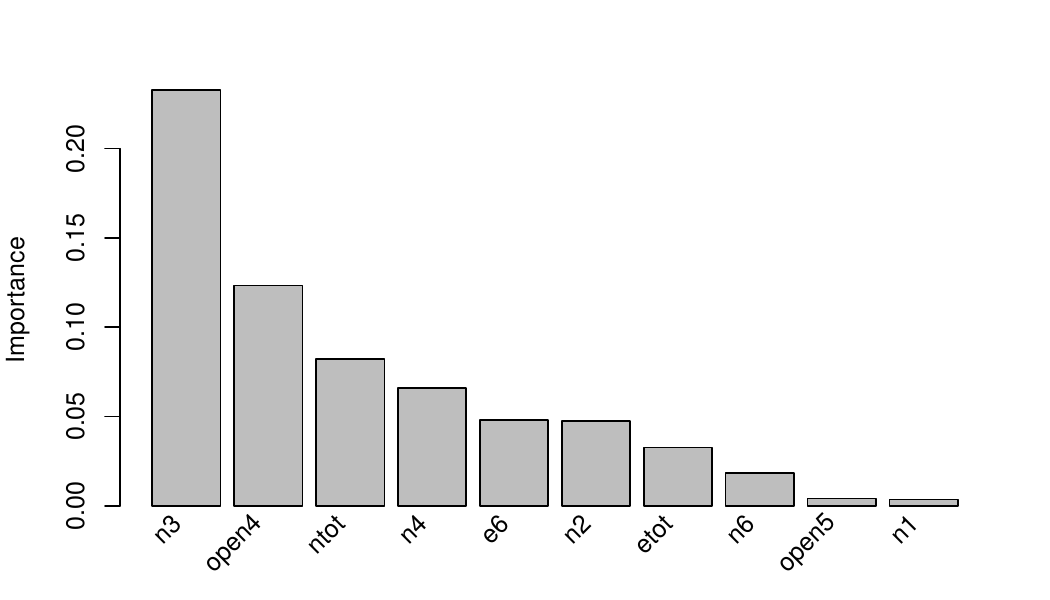}
\caption{Input variable importances for the prediction rule ensemble for predicting depression.}
\label{fig:importance}
\end{figure}

Figure~{\ref{fig:importance}} displays the
input variable importances. The two most important input variables for
predicting depressive symptoms are a neuroticism (\code{n3}) and an
openness subscale (\code{open4}). In addition to a plot of input
variable importances, \code{importance} returns a list of base learner
and variable importances, respectively (i.e., \code{baseimps} and
\code{varimps}; only the first three rows of the former are shown
here):
\begin{Schunk}
\begin{Sinput}
R> imps$baseimps[1:3, ]
\end{Sinput}
\begin{Soutput}
    rule           description    imp coefficient     sd
1     n3     2 <= n3 <= 30.225 0.1476      0.1752 6.6030
2 rule80 n4 > 15 & open4 <= 13 0.1281      2.4010 0.4183
3 rule88 ntot <= 109 & e6 > 15 0.0806     -1.3568 0.4656
\end{Soutput}
\end{Schunk}
We can plot (a subset of) the final ensemble using the \code{plot}
function. Below, \code{standardize = TRUE} is specified so that the
importances in the plots are standardized, \code{nterms = 6} so that
only the six most important base learners will be plotted,
\code{plot.dim = c(2, 3)} so that the rules will be plotted in two rows
and three columns and \code{cex = 0.7} to scale the size of node and
edge labels to fit the plot size:
\begin{Schunk}
\begin{Sinput}
R> plot(carrillo.ens, nterms = 6, plot.dim = c(2, 3), standardize = TRUE, 
+    cex = 0.7)
\end{Sinput}
\end{Schunk}
\begin{figure}[t!]
\centering
\includegraphics[width = 0.85\textwidth, trim = 0 20 0 15, clip]{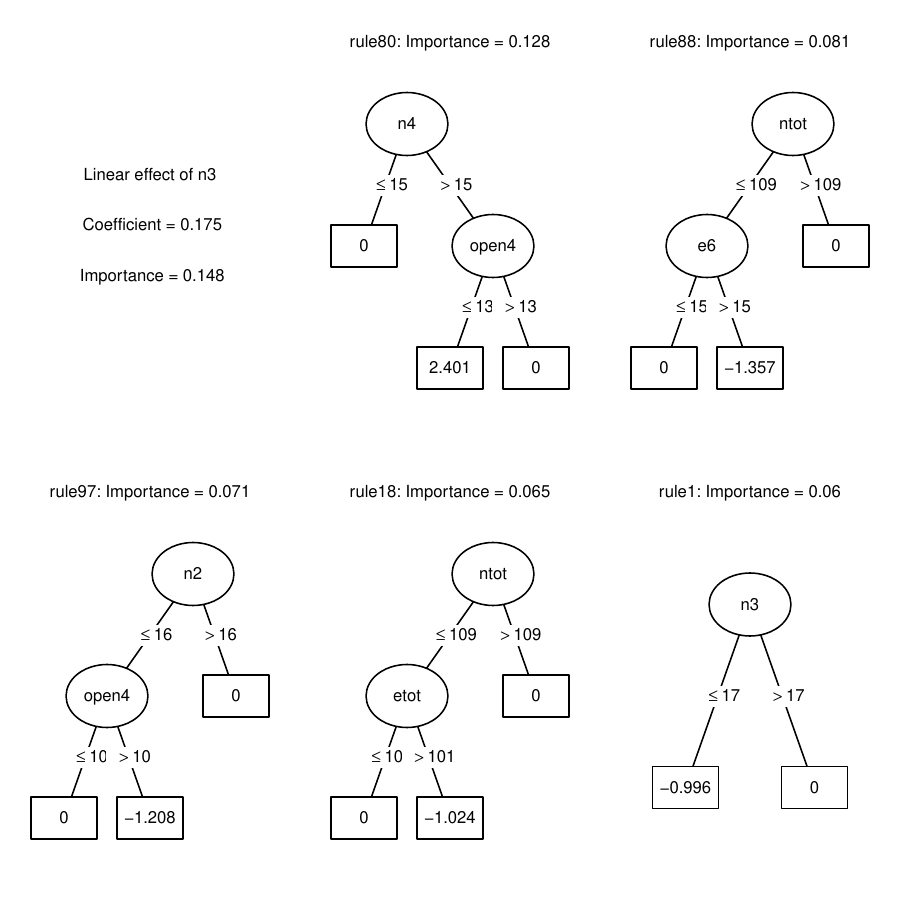}
\caption{The six most important base learners in the prediction rule ensemble for predicting depression.}
\label{fig:ensemble}
\end{figure}
  
Figure~{\ref{fig:ensemble}} displays the six most important base
learners in the ensemble. The most important base learner is a linear
term, \code{n3}. The second most important base learner is a rule
involving \code{n4} and \code{ open4}. Together, these base learners
indicate a positive association of depressive symptomtomatology with
neuroticism, and a negative association with extraversion and
openness.
  
To further inspect the shape of the effect of individual input variables, we can obtain a partial dependence plot using the \code{singleplot} function:
    
\begin{Schunk}
\begin{Sinput}
R> singleplot(carrillo.ens, varname = "ntot")
\end{Sinput}
\end{Schunk}
\begin{figure}[t!]
\centering
\includegraphics[width=0.75\textwidth, trim = 0 10 0 10, clip]{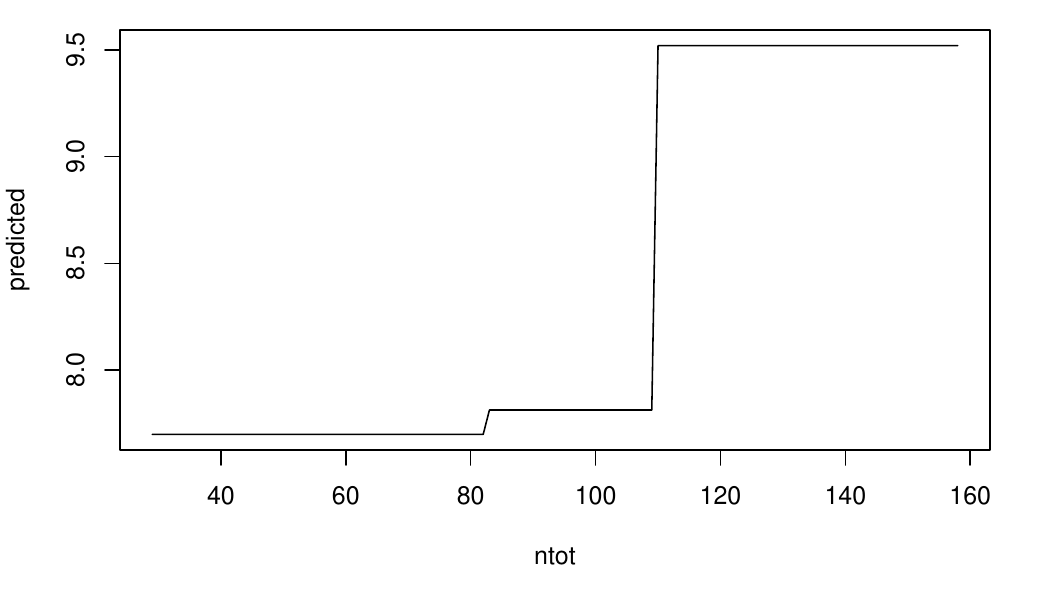}
\caption{Partial dependence plot of the response variable on input variable \code{ntot}.}
\label{fig:singleplot}
\end{figure}
  
Figure~\ref{fig:singleplot} displays the
partial dependence plot, which indicates a monotonically increasing,
rather stepwise association between the neuroticism scale and
depressive symptomatology. To inspect the combined association between
a pair of predictor variables and the response, we can employ the
\code{pairplot} function:
\begin{Schunk}
\begin{Sinput}
R> pairplot(carrillo.ens, varnames = c("n4", "open4"),
+    col = grey(seq(1, 0.4, by = -0.01)))
\end{Sinput}
\end{Schunk}
\begin{figure}[t!]
\centering
\includegraphics[width=0.6\textwidth, trim = 0 10 0 10, clip]{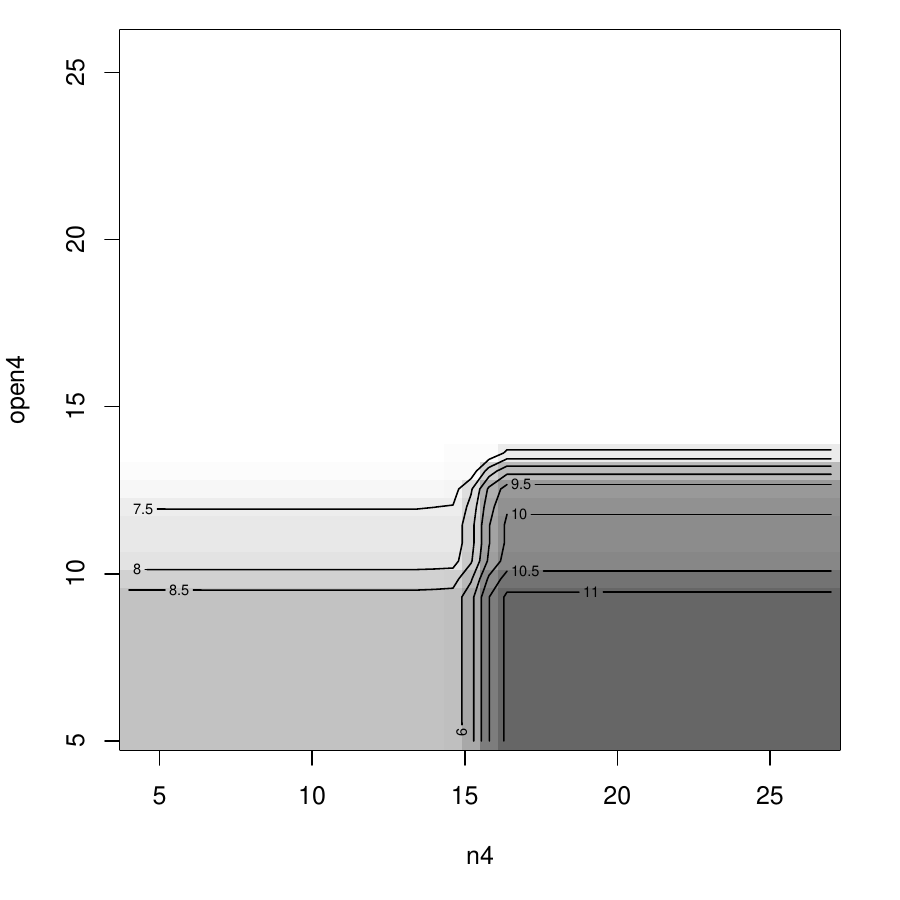}
\caption{Partial dependence plot of the response variable on input variables \code{n4} and \code{open4}.}
\label{fig:pairplot}
\end{figure}
  
By default, \code{pairplot} employs a plotting color sequence going
from yellow (lower values) to red (higher values). Here, a sequence
from light to dark grey has been specified through the \code{col}
argument. Figure~{\ref{fig:pairplot}} displays the partial dependence
of the depression variable on \code{n4} and \code{open4}. The plot
indicates that depressive symptomatology increases with increasing
values of the neuroticism subscale, and decreases with increasing
values of the openness subscale.

The pattern revealed by the partial dependence may reflect an interaction, or two main effects. If we want to assess and test the presence of interaction effects, we can employ the \code{interact} and \code{bsnullinteract} functions. The latter fits PREs on bootstrapped null-interaction datasets, that is, bootstrap sampled datasets from which interactions are known to be absent. Generating these null-interaction models is computationally intensive, therefore \code{bsnullinteract} generates ten null-interaction models, by default. To obtain a more precise estimate of the null distribution of interaction test statistics, we increase the number of generated null-interaction ensembles by specifying the \code{nsamp} argument (which will take substantially longer to compute). Because generating the null-interaction ensembles requires random sampling and permutation of the training data, we first set the random seed:
    
\begin{Schunk}
\begin{Sinput}
R> set.seed(43)
R> nullmods <- bsnullinteract(carrillo.ens, nsamp = 100)
\end{Sinput}
\end{Schunk}
Next, we obtain interaction test statistics for both the fitted and null-interaction models:
    
\begin{Schunk}
\begin{Sinput}
R> int.carrillo <- interact(carrillo.ens, nullmods = nullmods, 
+    varnames = c("n4", "open4", "e6"))
\end{Sinput}
\end{Schunk}
\begin{figure}[t!]
\centering
\includegraphics[width=0.5\textwidth, trim = 0 12 0 20, clip]{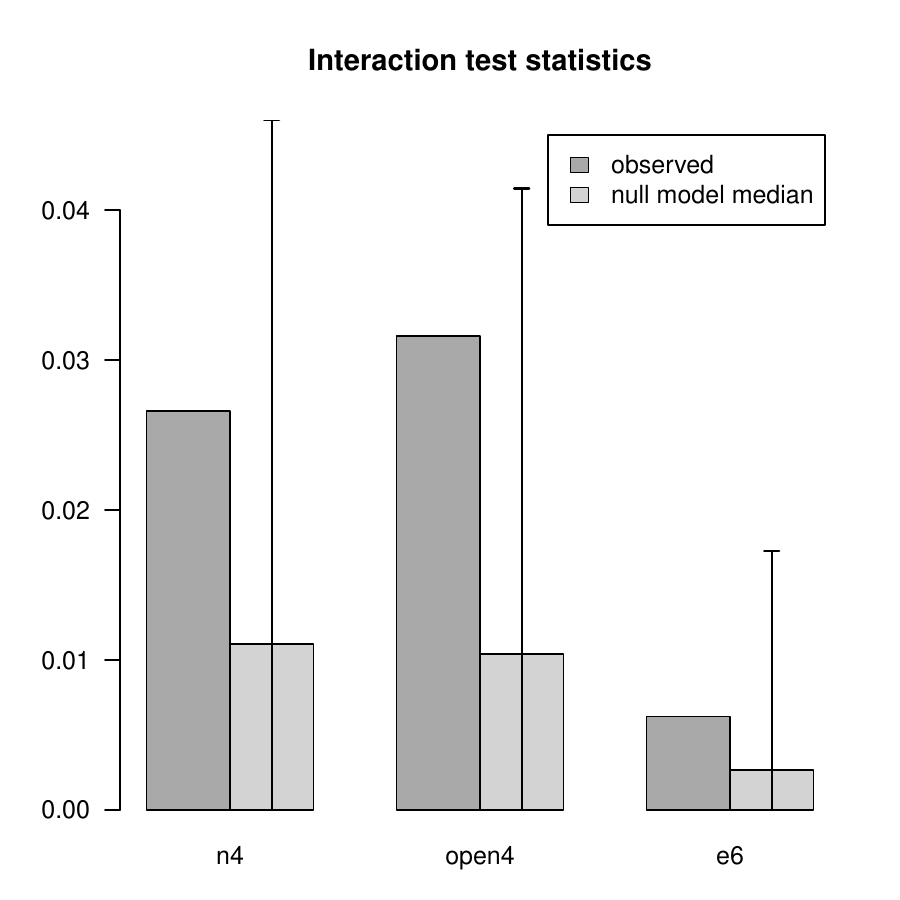}  
\caption{Interaction test statistics for input variables \code{n4},
  \code{open4} and \code{e6}. The darker grey bars represent
  interaction strengths of the fitted ensemble; the lighter grey bars
  represent the median interaction strength in the null interaction
  models, with the error bars indicating the 0.05 and 0.95 quantiles
  of the distribution.}
\label{fig:interact}
\end{figure}
  
Figure~{\ref{fig:interact}} displays the interaction test statistics
for \code{n4}, \code{open4} and \code{e6}. The darker bars represent
the observed interaction strengths in the fitted ensemble. The lighter
bars represent the median interaction strength in the null-interaction
models, with the error bars indicating the 0.05 and 0.95 quantiles of
the distribution. The plot indicates that none of the three specified
predictor variables are involved in interactions.

Finally, using the \code{cvpre} function, we estimate the
out-of-sample prediction error of the fitted ensemble through full
$k$-fold cross-validation. By default, the number of folds is set to
ten. As cross-validation involves random sampling of observations, we
first set the random seed:
    
\begin{Schunk}
\begin{Sinput}
R> set.seed(44)
R> cv.carrillo <- cvpre(carrillo.ens)
\end{Sinput}
\begin{Soutput}
$MSE
      MSE        se 
47.025425  8.065651 

$MAE
      MAE        se 
5.1876512 0.4256812 
\end{Soutput}
\end{Schunk}
The printed results show the mean squared error (MSE) and mean
absolute error (MAE) and their respective standard errors. For
classification, mean squared and mean absolute error loss would be
returned. With the code above, these accuracy estimates are also saved
in
\code{cv.carrillo$accuracy}. In addition, \code{cv.carrillo$cvpreds}
contains the cross-validated predictions, which can be used to
calculate alternative indices of predictive accuracy.

\subsection{Customizing rule induction and model selection}
\label{subsec:custom}

The default settings of \code{pre} employ a sub-sampling (i.e.,
\code{sampfrac = 0.5}) and boosting (i.e., \code{learnrate = 0.01})
approach. As in most tree boosting algorithms, maximum tree depth is
also limited by default (i.e., \code{maxdepth = 3L}). Note that, in
addition to the maximum tree depth, the \code{ctree} function
simultaneously employs an additional stopping criterion: by default an
$\alpha$ level of 0.05 for split selection is used. That is, no split
is performed when all of the partitioning variables show $p$ values
$>0.05$. The latter can be adjusted through specification of the
\code{tree.control} argument.

The following examples aim to illustrate how several well-known tree
ensemble approaches can be mimicked. The resulting ensembles are not
shown here, but will likely be more complex than ensembles generated
with the default settings.

\subsubsection{Bagging}

A bagging approach can be mimicked by setting \code{sampfrac = 1} (to employ bootstrap sampling), specifying a learning rate of 0 and applying no a-priori restrictions on tree size (i.e., employing unpruned trees):  
\begin{Schunk}
\begin{Sinput}
R> set.seed(42)
R> car.ens.bag <- pre(bdi ~ ., data = carrillo, sampfrac = 1, 
+    maxdepth = Inf, learnrate = 0, 
+    tree.control = partykit::ctree_control(alpha = 1))
\end{Sinput}
\end{Schunk}
Note that increasing the value of \code{maxdepth} will increase both computation time and complexity of the final ensemble.

\subsubsection{Random forest}

A random forest approach can be mimicked by additionally restricting the number of predictor variables considered for selecting each split through the \code{mtry} argument:
\begin{Schunk}
\begin{Sinput}
R> set.seed(42)
R> car.ens.ranfor <- pre(bdi ~ ., data = carrillo, sampfrac = 1, 
+    maxdepth = Inf, learnrate = 0, 
+    tree.control = partykit::ctree_control(alpha = 1),
+    mtry = ceiling(sqrt(ncol(carrillo))))
\end{Sinput}
\end{Schunk}
Additionally setting \code{tree.unbiased = FALSE} would employ the \code{rpart} implementation of the CART algorithm, which would most closely resemble the original bagging and random forest approaches. Note that this will generally yield more complex ensembles.

\subsubsection{Original RuleFit approach}

The default settings of the original RuleFit implementation can be
mimicked by employing the CART algorithm, and letting the sampling
fraction and number of cross-validation folds depend on the number of
effective observations, which is equal to the sample size for
regression. Furthermore, the maximum tree depth should be determined
by a random number generating function, allowing for varying tree
depths \citep{FrieyPope08, FrieyPope12}. Such a function can be
generated using the \code{maxdepth_sampler} function, which by default
samples from a distribution with an average maximum tree depth of two
(as in the original RuleFit implementation):
\begin{Schunk}
\begin{Sinput}
R> neff <- nrow(carrillo) 
R> set.seed(42)
R> car.ens.rulef <- pre(bdi ~ ., data = carrillo, tree.unbiased = FALSE,
+    maxdepth = maxdepth_sampler(), 
+    sampfrac = min(1, (11 * sqrt(neff) + 1) / neff),
+    nfolds = round(min(20, max(0, 5200 / neff - 2))))
\end{Sinput}
\end{Schunk}
Furthermore, to fully mimic the original RuleFit approach, the minimum cross-validated error criterion should be employed to obtain the final ensemble:
\begin{Schunk}
\begin{Sinput}
R> print(car.ens.rulef, penalty.par.val = "lambda.min")
\end{Sinput}
\end{Schunk}
\subsubsection{Regularized single tree}

An extremely sparse PRE would take the nodes of only a single tree as
the initial ensemble. The predictions of the resulting ensemble would
represent regularized node means of the single (pruned) tree. When
only a single tree is grown, bootstrap or sub-sampling may not improve
predictive accuracy, so we first define a custom sampling function,
which returns all indices of the training observations:
\begin{Schunk}
\begin{Sinput}
R> samp_func <- function(...) 1:nrow(carrillo)
\end{Sinput}
\end{Schunk}
Then we specify only a single tree to be grown and no linear terms to be included:
\begin{Schunk}
\begin{Sinput}
R> car.ens.tree <- pre(bdi ~ ., data = carrillo, ntrees = 1, 
+    type = "rules", sampfrac = samp_func)
\end{Sinput}
\end{Schunk}
\subsubsection{Regularization of the final ensemble}

Through the ellipsis (\code{...}) \code{pre} allows for passing
additional arguments to the \code{cv.glmnet} function. For example,
this allows for specifying the elastic net mixing parameter
$\alpha$. By default, the lasso penalty is employed (i.e., \code{alpha
  = 1}), but specifying \code{alpha = 0} would yield the ridge
penalty, and values $0 < \alpha < 1$ would yield the elastic net
penalty. An unpenalized solution can be obtained by supplying a
pre-specified range for the penalty parameter $\lambda$, including
0. Such an approach will likely yield sub-optimal sparsity and
accuracy, but may in rare cases be preferred over a penalized
solution. Note that the \code{cv.glmnet} function only permits
specification of multiple $\lambda$ values, so we have to specify at
least two $\lambda$ values:
\begin{Schunk}
\begin{Sinput}
R> set.seed(42)
R> car.ens.unp <- pre(bdi ~ ., data = carrillo, lambda = c(0, 1))
\end{Sinput}
\end{Schunk}
To obtain the unpenalized solution, all methods and functions applied to the resulting object need to specify \code{penalty.par.val = 0}: 
\begin{Schunk}
\begin{Sinput}
R> print(car.ens.unp, penalty.par.val = 0)
\end{Sinput}
\end{Schunk}
Finally, to employ a different loss function than the default
squared-error or log-likelihood criterion, \code{type.measure} can be
set to, for example, \code{"class"} (for misclassification error) or
\code{"mae"} (for mean absolute error):
\begin{Schunk}
\begin{Sinput}
R> set.seed(42)
R> car.ens.mae <- pre(bdi ~ ., data = carrillo, type.measure = "mae")
\end{Sinput}
\end{Schunk}
\section{Empirical evaluation}
\label{sec:Empirical} 

\subsection{Method}

\subsubsection{Datasets}

Four benchmark datasets were employed to compare the methods: two
regression datasets (\code{BostonHousing} and \code{Ozone}) and two
classification datasets (\code{BreastCancer} and \code{Ionosphere}), each obtained
from the UCI Machine Learning Repository \citep{DuayGraf19} through the
\pkg{mlbench} package \citep[version 2.1-1;][]{LeisyDimi12}. Only
complete observations were included in the analyses. Specifically, no
observations were removed from the \code{BostonHousing} dataset. From
the \code{BreastCancer} dataset, 16 cases were removed due to missing
values. From the \code{Ionosphere} dataset, one variable was removed due to
zero variance. From the \code{Ozone} dataset, a categorical variable
with a large number of categories was removed to reduce computation
time (i.e., day of month), one variable was removed due to a large
number of missing values (temperature at El Monte, California) and a
total of 36 cases were removed due to missing
values. Table~{\ref{tab:datasets}} provides the resulting total sample
sizes and numbers of predictor variables for each dataset.

\begin{table}[t!]
\centering
\begin{tabular}{llcc}
\hline
Dataset         & Response        & $p$ & $N$ \\
\hline
\code{BostonHousing}   & numeric         & 13  & 506 \\
\code{BreastCancer}    & binary factor   & 9  & 683 \\
\code{Ionosphere}      & binary factor   & 33  & 351 \\
\code{Ozone}           & numeric         & 10  & 330 \\
\hline
\end{tabular}
\caption{Benchmark datasets used for comparing performance; $p$ refers to the total number of predictor variables in the dataset, $N$ refers to the total sample size.}
\label{tab:datasets}
\end{table}

\subsubsection{Model fitting procedures}

The performance of \code{pre} was compared with that of random
forests, single regression trees, lasso penalized linear regression
and the original RuleFit implementation. All analyses were performed
in \proglang{R} \citep[version 3.6.1;][]{R19}. To fit PREs, function
\code{pre} from package \pkg{pre} \citep[version
0.7.2;][]{FokkyChri19} was employed. In addition, to fit PREs with the
original RuleFit implementation, \pkg{RuleFit} version 3 \citep{FrieyPope12}
for Windows was obtained from
\url{https://statweb.stanford.edu/~jhf/R_RuleFit.html} and function
\code{rulefit} was employed. To fit single trees, function
\code{ctree} from package \pkg{partykit} \citep[version
1.2-6;][]{HothyZeil15} was used to fit conditional inference trees,
and function \code{rpart} from package \pkg{rpart} \citep[version
4.1-15;][]{TheryAtki19} was used to fit CART trees. To fit random
forests, function \code{cforest} from package \pkg{partykit} was used
to fit random forests based on conditional inference trees, and
function \code{randomForest} from package \pkg{randomForest}
\citep[version 4.6-14;][]{LiawyWien02} was used to fit random forests
based on CART trees. To fit lasso penalized linear regression models,
function \code{cv.glmnet} from package \pkg{glmnet} \citep[version
3.0-2;][]{FrieyHast10} was used.

% The performance of \code{pre} was compared with that of random
% forests, single regression trees, lasso penalized linear regression
% and the original RuleFit implementation. All analyses were performed
% in \proglang{R} \citep[version 3.6.1;][]{R17}. To fit PREs, function
% \code{pre} from package \pkg{pre} \citep[version
% 0.7.2;][]{FokkyChri17} was employed. In addition, to fit PREs with the
% original RuleFit implementation, \pkg{RuleFit} version 3
% \citep{FrieyPope12} was obtained from
% \url{https://statweb.stanford.edu/~jhf/R_RuleFit.html} and function
% \code{rulefit} was employed. To fit single trees, function
% \code{ctree} from package \pkg{partykit} \citep[version
% 1.2.6;][]{HothyZeil15} was used to fit single conditional inference
% trees, and function \code{rpart} from package \pkg{rpart}
% \citep[version 4.1-15;][]{TheryAtki17} was used to fit single CART
% trees. To fit random forests, function \code{cforest} from package
% \pkg{partykit} was used to fit random forests based on conditional
% inference trees, and function \code{randomForest} from package
% \pkg{randomForest} \citep[version 4.6-14;][]{LiawyWien02} was used to
% fit random forests based on CART trees. To fit lasso penalized linear
% regression models, function \code{cv.glmnet} from package \pkg{glmnet}
% \citep[version 3.0.2;][]{FrieyHast10} was used.

All analyses employed default settings, with two exceptions: For
fitting PREs, maximum rule length was set to 4, instead of the default
value of 3. Furthermore, trees fitted with \pkg{rpart} were pruned
using the 1 standard error criterion. Note that careful tuning of
parameter settings using cross-validation approaches would likely
yield higher predictive accuracy for all methods.

For functions \code{rulefit} and \code{pre}, selection of the final
ensembles was performed using two different criteria. For each, a
final ensemble was obtained through lasso regression, with the value
of $\lambda$ set to minimize squared error loss based on
cross-validation. In addition, for \code{pre}, a sparser ensemble was
obtained through employing a $\lambda$ value yielding squared error
loss within one standard error above the minimum (which is the default
in package \pkg{pre}). For \code{rulefit}, a sparser ensemble was
obtained through forward stepwise regression for numeric outcomes,
and through forward stagewise regression with variable entry order
determined by lasso regression for binary outcomes. For
\code{pre}, the number of cross-validation replications was set to the
default value of 10; for \code{rulefit}, the default was also
employed, which is a function of sample size, yielding 6 to 14
replications in the current analyses.

\subsubsection{Assessment of performance}

To assess performance, the bootstrap cross-validation design for
benchmark experiments with real-world data of \cite{HothyLeis05} was
employed. From each dataset, 250 bootstrap samples were drawn. Each
bootstrap sample was used for training and predictive accuracy was
subsequently assessed using the test observations that were not
included in the bootstrap sampled training data. Predictive accuracy
was quantified through calculating mean squared error (MSE) in
regression problems and the area under the receiver operating curve
(AUC) value in classification problems. Interpretability was assessed
through counting the number of predictor variables, for the
\code{BostonHousing} and \code{Ionosphere} data. As it was assumed the
two random forest methods and the two single tree methods would yield
similar complexities, respectively, only the number of variables for
\code{cforest} and \code{ctree} were counted. For the
\code{BreastCancer} and \code{Ozone} datasets, the total number of
base learners with non-zero coefficients (i.e., the number of terms)
were counted.  The number of terms were only evaluated for \code{pre}
and \code{rulefit}. Finally, computation time in seconds was recorded
for every fitted model.

% latex table generated in R 3.4.3 by xtable 1.8-2 package
% Fri Oct 26 10:51:42 2018
\begin{table}[!t]
\centering
\begin{tabular}{lccccc}
  \hline
 & \code{BostonHousing} & \code{BreastCancer} & \code{Ionosphere} & \code{Ozone} & Mean rank \\ 
  \hline
randomForest & 11.75 (3.41) & 0.993 (0.003) & 0.979 (0.011) & 17.12 (2.22) & 1.25 \\ 
  cforest & 16.47 (4.46) & 0.993 (0.004) & 0.971 (0.014) & 17.78 (2.37) & 3.50 \\ 
  pre\_1se & 14.06 (3.69) & 0.991 (0.005) & 0.961 (0.018) & 17.72 (2.67) & 3.50 \\ 
  rulefit\_min & 12.90 (3.29) & 0.990 (0.006) & 0.961 (0.017) & 20.19 (3.02) & 4.00 \\ 
  pre\_min & 13.35 (3.48) & 0.991 (0.005) & 0.960 (0.019) & 18.85 (2.87) & 4.25 \\ 
  lasso & 27.58 (5.12) & 0.995 (0.002) & 0.902 (0.032) & 20.31 (2.27) & 5.75 \\ 
  ctree & 21.68 (4.94) & 0.975 (0.011) & 0.908 (0.032) & 24.88 (4.16) & 7.00 \\ 
  rulefit\_FS & 15.35 (4.46) & 0.945 (0.016) & 0.883 (0.032) & 24.32 (4.22) & 7.50 \\ 
  rpart & 25.34 (5.74) & 0.948 (0.018) & 0.892 (0.037) & 26.01 (4.07) & 8.25 \\ 
   \hline
\end{tabular}
\caption{Predictive accuracy for all methods and datasets. Values
  represent averages over 250 bootstrap samples, with standard
  deviations in parentheses. For the \code{BostonHousing} and
  \code{Ozone} data, mean squared error (MSE) was calculated; for the
  \code{BreastCancer} and \code{Ionosphere} data, the area under the receiver
  operating characteristic curve (AUC) values were calculated. Mean
  rank represents the ranking of the algorithms from highest to lowest
  predictive accuracy, averaged over the four datasets.}
\label{tab:acc}
\end{table}
\subsection{Results}

Table~\ref{tab:acc} presents average predictive
accuracies in each of the benchmark datasets. Overall, function
\code{randomForest} ranked highest in terms of predictive accuracy,
followed by \code{cforest} and \code{pre} (employing the 1 
standard-error criterion), followed by \code{rulefit}, followed by 
\code{pre} (each of the latter two employing the minimum cross-validation
error criterion). Overall, \code{pre} showed higher accuracy than
\code{rulefit}. Of note, \code{rulefit} employing the minimum cross
validation error criterion showed substantially better accuracy than
\code{rulefit} employing forward selection. The latter model showed
predictive accuracy comparable to single trees.

\begin{figure}[t!]
  \centering
\includegraphics[width=0.48\textwidth, trim = 0 0 0 15, clip]{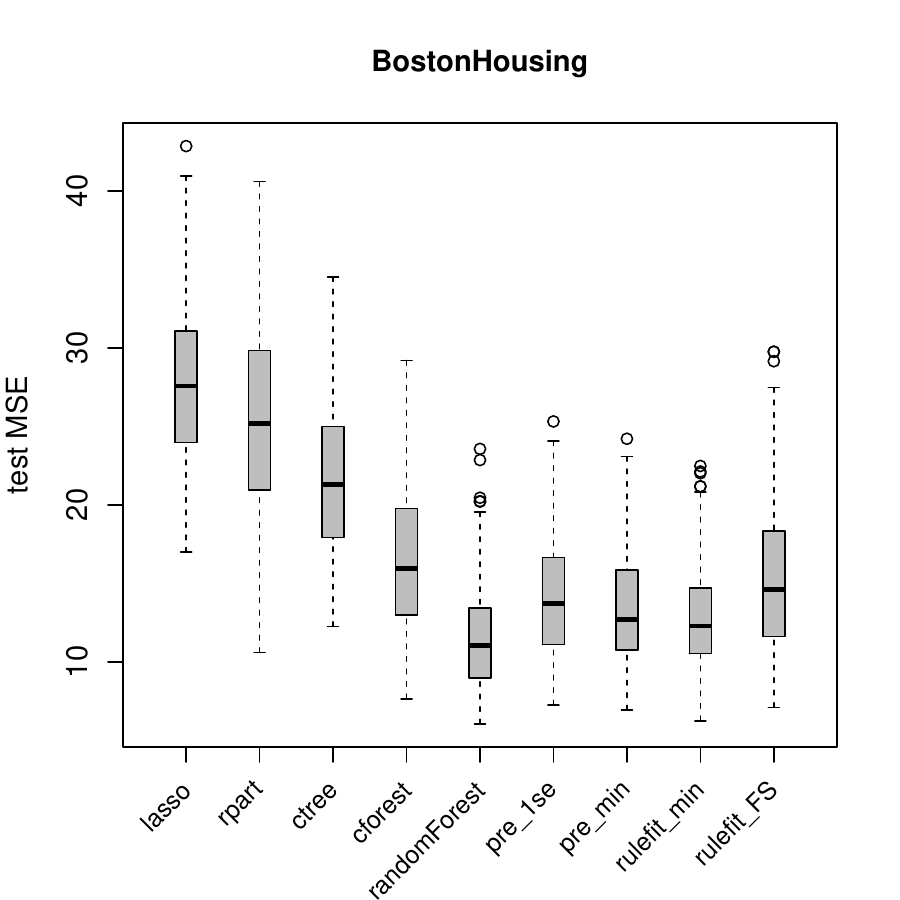}
\includegraphics[width=0.48\textwidth, trim = 0 0 0 15, clip]{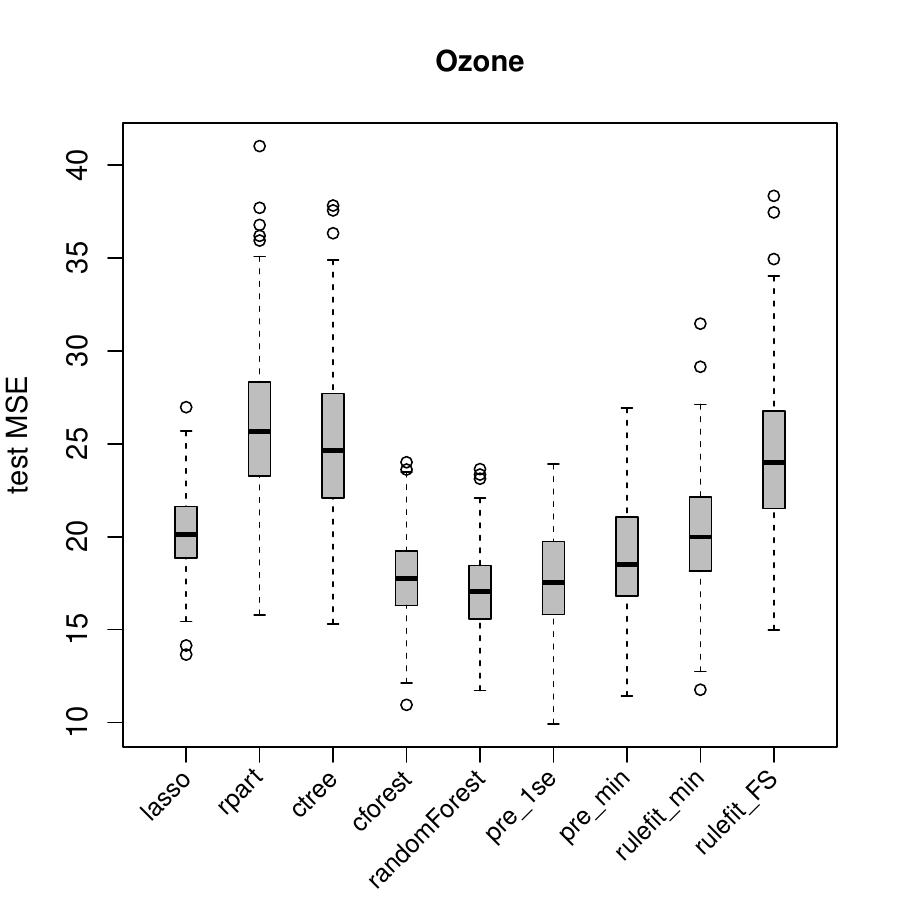}

\includegraphics[width=0.48\textwidth, trim = 0 0 0 15, clip]{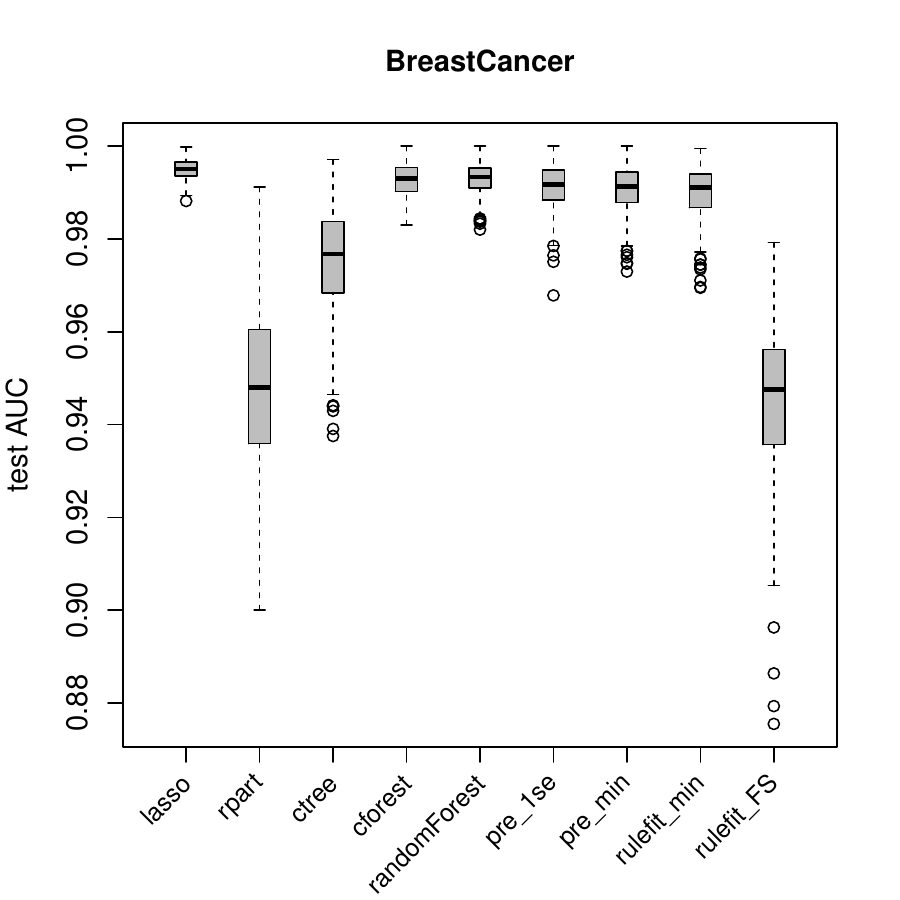}
\includegraphics[width=0.48\textwidth, trim = 0 0 0 15, clip]{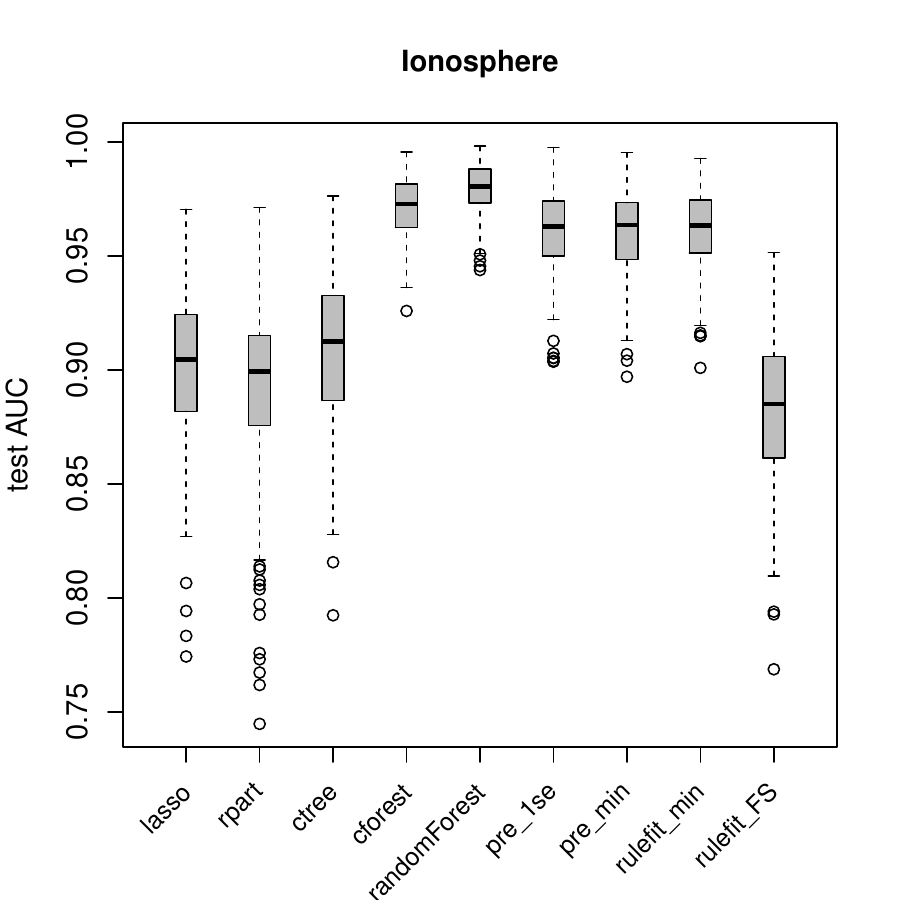}
\caption{Predictive accuracy for the different algorithms in 250
  bootstrap samples of the benchmark datasets. Three AUC values for
  \pkg{rpart} were winsorized to 0.90 in the \code{BreastCancer}
  dataset to improve the graphical presentation. MSE = mean squared
  error; AUC = area under the receiver operating characteristic curve;
  FS = forward selection.}
\label{fig:accuracy}
\end{figure}

\begin{figure}[t!]
  \centering
\includegraphics[width=0.48\textwidth, trim = 0 0 0 15, clip]{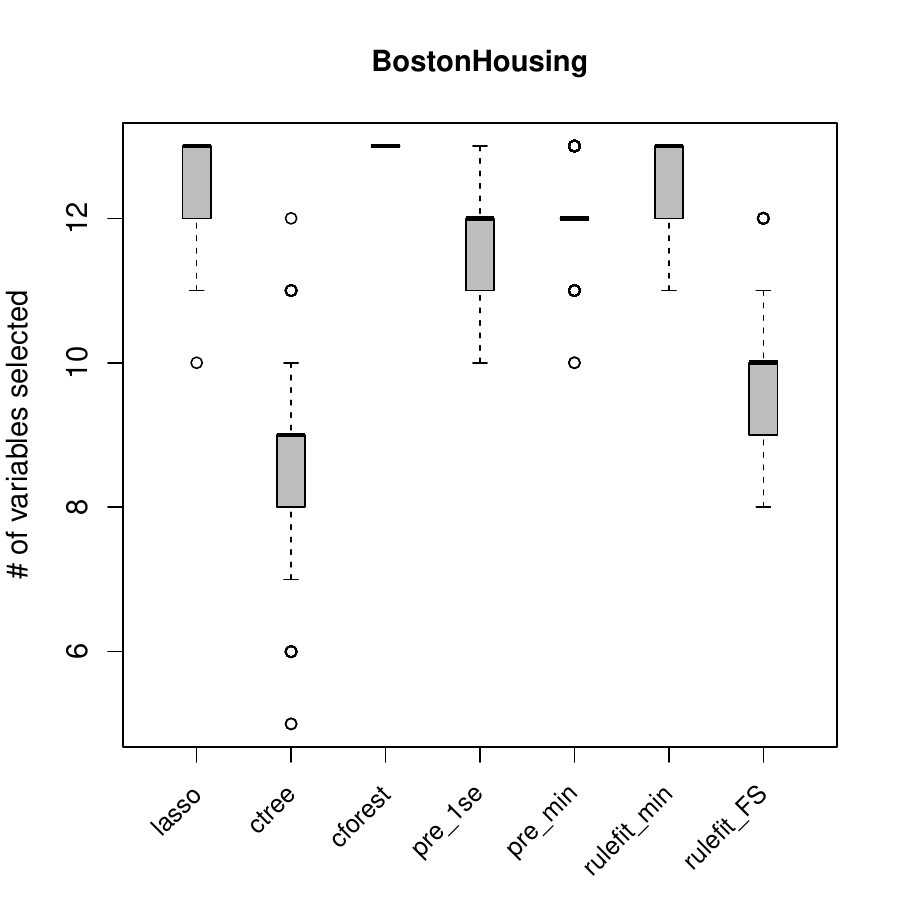}
\includegraphics[width=0.48\textwidth, trim = 0 0 0 15, clip]{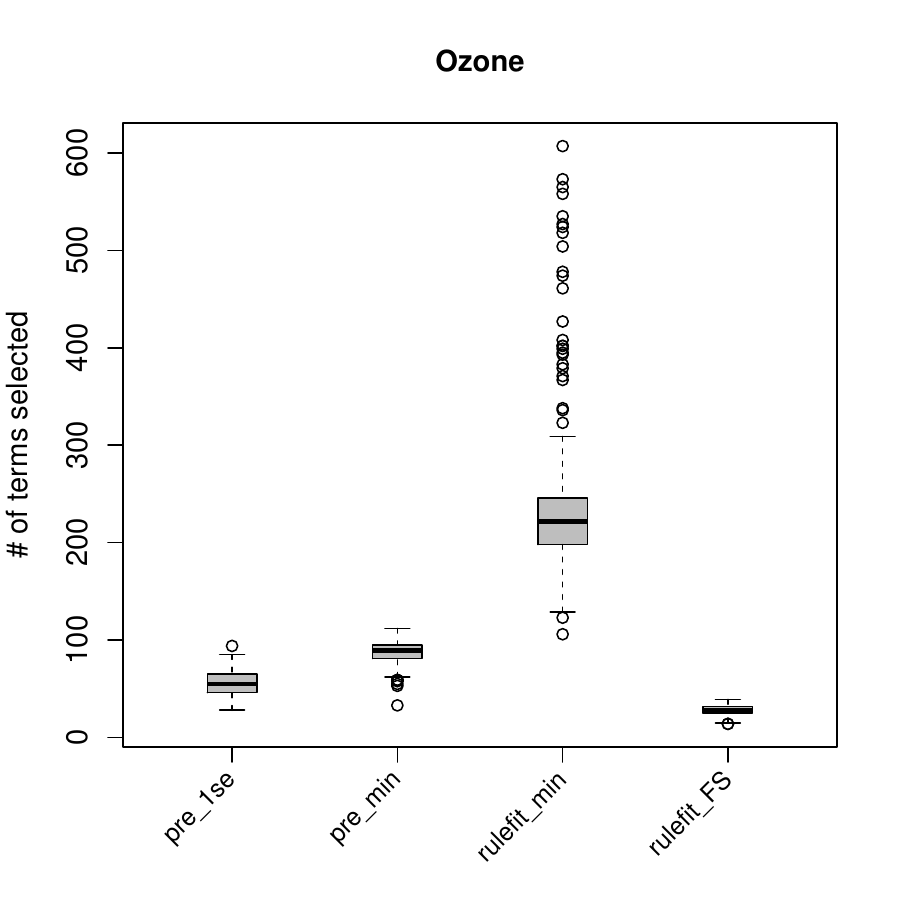}

\includegraphics[width=0.48\textwidth, trim = 0 0 0 15, clip]{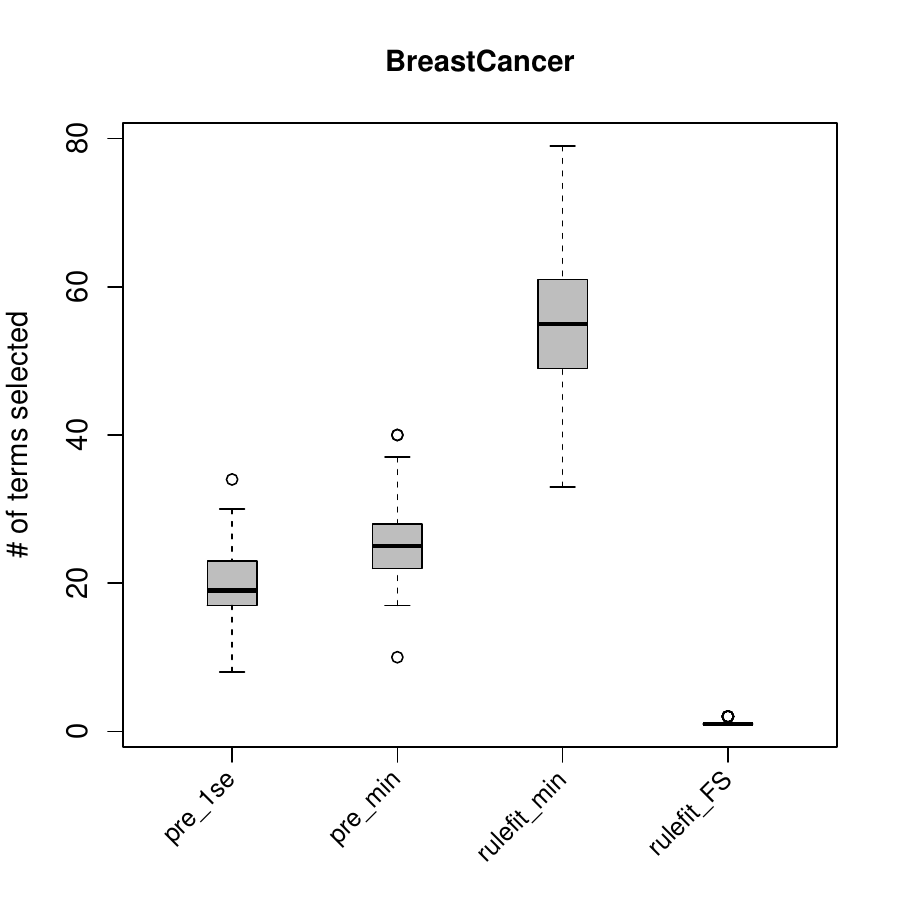}
\includegraphics[width=0.48\textwidth, trim = 0 0 0 15, clip]{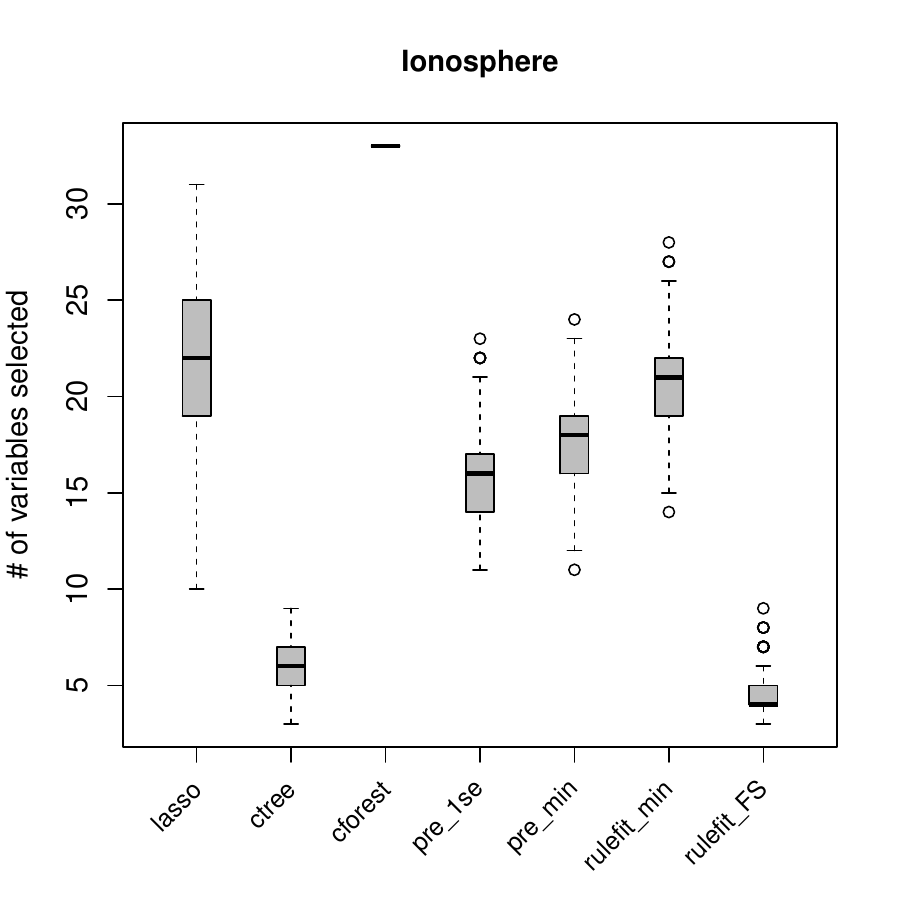}
\caption{Distribution of the number of predictor variables or number
  of terms selected in 250 bootstrap samples of the benchmark
  datasets.}
\label{fig:interp}
\end{figure}

Figure~\ref{fig:accuracy} depicts the
distributions of predictive accuracy across the four benchmark
datasets. The boxplots indicate that the most accurate methods also
show the least variation in predictive accuracy.
Figure~\ref{fig:interp} depicts the fitted
models' complexities. The random forests always included all predictor
variables, providing the least sparse solutions. After random forests,
lasso regression provided the least sparse solutions, followed by PREs
selected using lasso regression. The sparsest solutions were provided
by singles trees (ctrees) and rulefit ensembles with forward stepwise
selection. Comparing the complexity of PRE methods yields a similar
pattern in all datasets: \code{rulefit} with forward selection
provided the sparsest PREs, followed by \code{pre} with the
$1-\text{SE}$ rule, followed by \code{pre} with minimum cross
validated error, followed by \code{rulefit} with minimum cross
validated error. Taken together with the findings on predictive
accuracy, these results indicate that \code{pre} may provide a better
trade-off between accuracy and interpretability than \code{rulefit}.

\begin{figure}[t!]
\centering
\includegraphics[width=0.48\textwidth, trim = 0 0 0 45, clip]{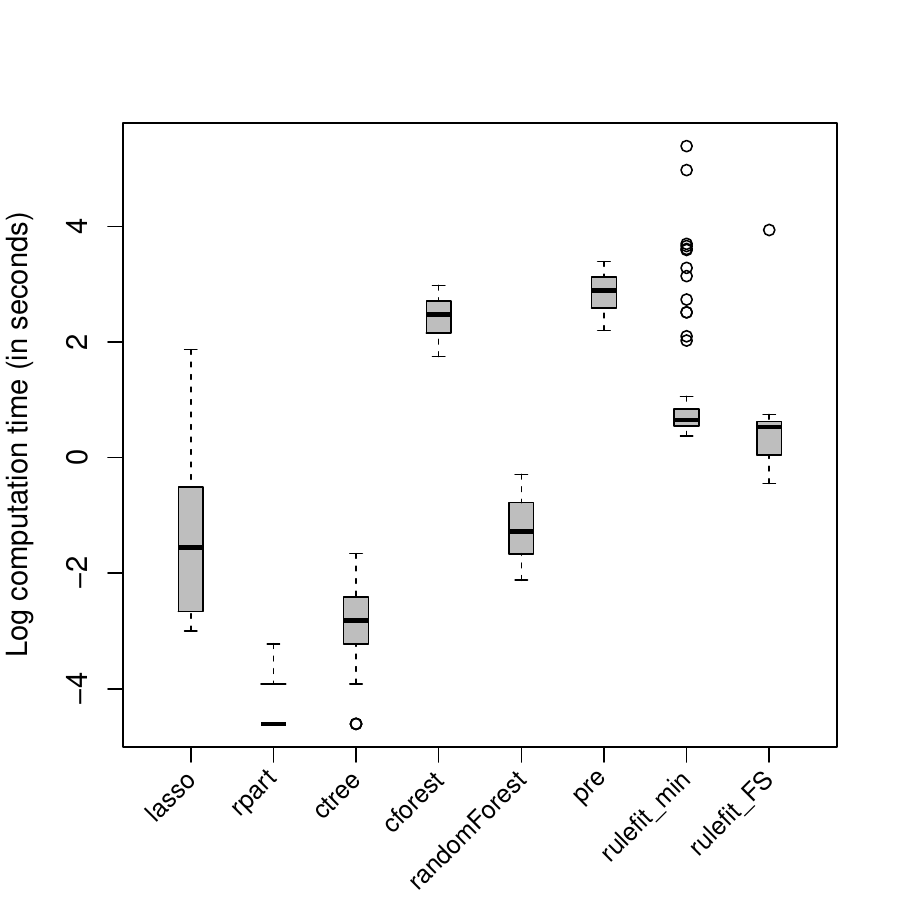}
\caption{Distribtuions of the log of computation time (in seconds) across all bootstrap sampled datasets.}
\label{fig:comp_time}
\end{figure}

Figure~{\ref{fig:comp_time}} depicts the computation time
distributions for all methods. Note that for \code{pre}, only a single
boxplot is depicted, because the $1-\text{SE}$ and minimum cross
validated error solutions are obtained from the same fitted
model. Figure~\ref{fig:comp_time} shows a clear computational
disadvantage for \code{pre}, with an average computation time of 18
seconds. Most computation time for \code{pre} is spent on the fitting
of ctrees, which is computationally more demanding than fitting CART
trees. This can also be observed in the computation time differences
between \code{ctree} (0.066 seconds, on average) and \code{rpart}
(0.013 seconds, on average), and between \code{cforest} (12.1 seconds,
on average) and \code{randomForest} (0.3 seconds, on average). The
\code{rulefit} function also employs CART trees and yielded average
computation times of 1.5 seconds (with forward selection) and 2.6
seconds (with lasso selection).

\section{Comparison with other packages}
\label{sec:Comparison}

A number of algorithms and software packages for fitting PREs have been developed over the last years. An extensive empirical comparison is outside the scope of the current paper, but several packages are discussed and compared below. Following \cite{FranyWitt98}, we distinguish between two strategies for generating rules: Indirectly, through transforming the nodes of one or more decision trees to a set of rules, and directly, through for example a sequential covering approach \citep{Furn99}.    

The indirect approach to rule generation is employed in the method of
\cite{FrieyPope08}. The \pkg{RuleFit} program \citep{FrieyPope12}, which is
written in \proglang{Fortran} and can be executed through an
\proglang{R} interface, provides a fast implementation of the method,
which was also observed in Section~\ref{sec:Empirical}. In addition,
several \proglang{R} packages implement a two-step approach similar to
that of \cite{FrieyPope08}: \pkg{inTrees} \citep{Deng14},
\pkg{horserule} \citep{NaleyVill17pkg} and \pkg{nodeHarvest}
\citep{Mein10} also generate rules from the nodes of tree ensembles,
after which weights or coefficients are estimated to construct a final
ensemble. For this second step, each package employs a different
approach: Package \pkg{inTrees} uses a sequential covering approach
\citep{Furn99} to select a sparse final ensemble of rules. Package
\pkg{horserule} estimates rule coefficients using a Bayesian linear
model with horseshoe prior \citep{CarvyPols10}. Although the horseshoe
estimator has been found to yield better predictive accuracy than
lasso regression, it does not enforce a sparse solution as it does not
shrink coefficients to a value of zero. Finally, package
\pkg{nodeHarvest} obtains node weights through solving a quadratic
programming problem with linear inequality constraints, yielding
non-zero weights for only a (small) subset of nodes \citep{Mein10}.

The main advantage of the node harvest approach is that it does not
require selection of a tuning parameter, as estimation with the lasso
does. Also, node harvest predictions are given by weighted node means,
which may aid interpretation: If an observation falls only into a
single node, the prediction is the average response among the training
observations in that node. In contrast, the coefficients in
Equation~{\ref{eq:predictive_model}} are shrunken towards zero and
cannot be interpreted as node means. On the other hand, the lasso
regression model in Equation~{\ref{eq:lasso}} can
more easily be extended to include linear (and other) functions of
predictor variables.

\pkg{C5.0} \citep{Quin93, Rule17} also employs an indirect approach to
rule learning. \pkg{C5.0} performs classification only, is written in
\proglang{C} and is available as a standalone executable
file. Alternatively, package \pkg{C50} \citep{Kuhn15} provides an
\proglang{R} interface. The predecessor \pkg{C4.5} is described in detail in
\citep{Quin93}, while the documentation on the implementation of
\pkg{C5.0} is limited \citep{Rule17}, but \cite{KuhnyJohn13} provide a
rather complete description. \pkg{C5.0} allows for generating PREs
from the nodes of a single tree or a boosted tree ensemble. In the
former case, a set of rules is derived from the nodes of a single tree
and simplified through pruning and deletion of rules, so as to
minimize prediction error. Predictions for new observations are
generated by a weighted majority vote of the rule ensemble. When
boosting is applied, observation weights are adjusted based on the
current classification error in every iteration. Predictions for new
observations are then given by the average of the predicted class
probability of each of the rule sets. Although \pkg{C5.0} tree
ensembles rank among the most accurate classifiers, \pkg{C5.0} rule
ensembles have been found to perform less well
\citep[e.g.,][]{FernyCern14}.

\pkg{Weka}'s \citep{HallyFran09} sub-package \pkg{classifiers.rules}
implements several algorithms for deriving PREs: \code{JRip}
\citep[implementing the RIPPER algorithm of][]{Cohe95},
\code{M5Rules} \citep{Quin92, HolmyHall99} and \code{PART}
\citep{FranyWitt98}. \code{M5Rules} builds a PRE for regression
through a sequential covering approach; it builds a tree in every
iteration and takes the best node as a rule. \code{PART} employs the
same approach for building a PRE for classification. \code{JRip}
generates rules directly through a sequential covering approach. As
the sequential covering approach is likely to be outperformed by
boosting \citep[e.g.,][]{CoheySing99, DembyKotl10}, these algorithms
will not be further discussed here.

Another algorithm that generates rules directly is ENDER
\citep{DembyKotl10}, which provides a very general framework for
generating boosted PREs. It is implemented in \pkg{RegENDER}
\citep{DembyKotl08}, which is written in \proglang{Java} and can be
executed from \pkg{Weka}. ENDER allows users to select from a range of
loss functions and regularization methods, thereby also encompassing
classification rule ensemble learners like SLIPPER \citep{CoheySing99}
and lightweight rule induction \citep{WeisyIndu00}. Notably,
\cite{DembyKotl10} report that predictive accuracy is hardly affected
by the choice of loss function, but substantially affected by the
regularization methods employed. They found regularization through
application of a learning rate, resampling of observations and
calculating coefficients on the complete training data instead of
sub-samples to yield improved accuracy of the final ensemble.

The main difference between \pkg{pre} and other packages employing an
indirect approach to rule generation is in the tree induction
algorithm. The results in Section~{\ref{sec:Empirical}} indicate that
conditional inference trees provide equal or better predictive
accuracy than CARTs. Also, \cite{SchayZeil07} found conditional
inference trees to yield lower complexity than C4.5 trees, but higher
complexity than CARTs. The current findings on the lower complexity of
\code{pre} compared to \code{rulefit} may be due to \code{ctree}'s
default stopping criterion, where splitting is halted when the null
hypothesis of independence between any of the input variables and the
response cannot be rejected in a node. The \code{rulefit} function
does not employ such a data-driven stopping criterion. The current
results suggest that \code{ctree}'s stopping criterion may improve both
sparsity and predictive accuracy of the final ensemble.

Finally, \pkg{pre} and {\pkg{RegENDER}} are similar in that they apply
regularization through boosting, sampling and global estimation of
rule coefficients. The main difference between the two packages lies
in the rule-generation approach employed. \cite{DembyKotl10} note that
the main advantage of their approach is that the rules are constructed
directly based on impurity measures. Specifying a minimum value for
improvement of the impurity measure then yields a natural stopping
criterion for building rules. However, the unbiased tree induction
algorithms employed by \pkg{pre} also provide a natural stopping
criterion, because the splitting is based on statistical testing. In
addition, the maximum number of conditions that may appear in rules,
as specified by the \code{maxdepth} argument of \code{pre}, provides
an additional stopping criterion for building rules.

\section{Conclusion}
\label{sec:Conclusion}

The current paper presented function \code{pre} from \proglang{R}
package \pkg{pre}, which allows for deriving prediction rule ensembles
for (multivariate) continuous, binary, multinomial, count and survival
outcomes. The fitting procedures and measures for interpretation as
implemented in package \pkg{pre} were discussed. Using an example
dataset on the prediction of depressive symptomatology, a rule
ensemble was derived and inspected. In four benchmark datasets, the
performance of function \code{pre} was compared with the original
RuleFit implementation, random forests, single trees and lasso
penalized regression models. Results indicated that \code{pre}
provided slightly better accuracy than the original RuleFit
implementation. Furthermore, \code{pre} provided accuracy
similar to random forests, while providing substantially lower
complexity than both random forests and RuleFit. The lower complexity
of \code{pre} is likely due to the use of unbiased recursive
partitioning methods, which do not have a preference for variables
with many possible splitting values and employ a statistical criterion
for split selection. Although \code{pre} provided a better trade-off
between complexity and accuracy, it also yielded the longest
computation times. Future developments will focus on reducing
computation time, improving modularity and extending the range of
methods that can be employed for selecting the final ensemble.

\section*{Acknowledgments}

The author would like to thank Benjamin Christoffersen for his
contributions to the development of package \pkg{pre}. Also, the author
would like to thank Henk Kelderman for commenting on an earlier
version of this manuscript. Part of the work in this publication was
carried out during a research visit to the University of Zurich,
supported through an International Short Visit grant from the Swiss
National Science Foundation (IZK0Z1\_175531).

\bibliography{ref}

\begin{thebibliography}{49}
\newcommand{\enquote}[1]{``#1''}
\providecommand{\natexlab}[1]{#1}
\providecommand{\url}[1]{\texttt{#1}}
\providecommand{\urlprefix}{URL }
\expandafter\ifx\csname urlstyle\endcsname\relax
  \providecommand{\doi}[1]{doi:\discretionary{}{}{}#1}\else
  \providecommand{\doi}{doi:\discretionary{}{}{}\begingroup
  \urlstyle{rm}\Url}\fi
\providecommand{\eprint}[2][]{\url{#2}}

\bibitem[{Beck \emph{et~al.}(1988)Beck, Steer, and Carbin}]{BeckyStee88}
Beck AT, Steer RA, Carbin MG (1988).
\newblock \enquote{Psychometric Properties of the Beck Depression Inventory:
  Twenty-Five Years of Evaluation.}
\newblock \emph{Clinical Psychology Review}, \textbf{8}(1), 77--100.
\newblock \doi{10.1016/0272-7358(88)90050-5}.

\bibitem[{Breiman(1996)}]{Brei96}
Breiman L (1996).
\newblock \enquote{Heuristics of Instability and Stabilization in Model
  Selection.}
\newblock \emph{The Annals of Statistics}, \textbf{24}(6), 2350--2383.
\newblock \doi{doi:10.1214/aos/1032181158}.

\bibitem[{Breiman \emph{et~al.}(1984)Breiman, Friedman, Olshen, and
  Stone}]{BreiyFrie84}
Breiman L, Friedman J, Olshen R, Stone C (1984).
\newblock \emph{Classification and Regression Trees}.
\newblock Wadsworth, New York, NY.

\bibitem[{Carrillo \emph{et~al.}(2001)Carrillo, Rojo, S\'{a}nchez-Bernardos,
  and Avia}]{CarryRojo01}
Carrillo J, Rojo N, S\'{a}nchez-Bernardos M, Avia M (2001).
\newblock \enquote{Openness to Experience and Depression.}
\newblock \emph{European Journal of Psychological Assessment}, \textbf{17}(2),
  130--136.
\newblock \doi{10.1027//1015-5759.17.2.130}.

\bibitem[{Carvalho \emph{et~al.}(2010)Carvalho, Polson, and
  Scott}]{CarvyPols10}
Carvalho CM, Polson NG, Scott JG (2010).
\newblock \enquote{The Horseshoe Estimator for Sparse Signals.}
\newblock \emph{Biometrika}, \textbf{97}(2), 465--480.

\bibitem[{Cohen(1995)}]{Cohe95}
Cohen WW (1995).
\newblock \enquote{Fast Effective Rule Induction.}
\newblock In \emph{Proceedings of the Twelfth International Conference on
  Machine Learning}, pp. 115--123. Morgan Kaufmann Publishers, San Mateo, CA.

\bibitem[{Cohen and Singer(1999)}]{CoheySing99}
Cohen WW, Singer Y (1999).
\newblock \enquote{A Simple, Fast, and Effective Rule Learner.}
\newblock In \emph{Proceedings of the National Conference on Artificial
  Intelligence}, pp. 335--342. John Wiley \& Sons Ltd, New York, NY.

\bibitem[{Costa and McCrae(1985)}]{CostyMcCr85}
Costa PT, McCrae RR (1985).
\newblock \emph{{The NEO Personality Inventory}}.
\newblock Psychological Assessment Resources, Odessa, FL.

\bibitem[{De~Bin \emph{et~al.}(2016)De~Bin, Janitza, Sauerbrei, and
  Boulesteix}]{DeBiyJani16}
De~Bin R, Janitza S, Sauerbrei W, Boulesteix AL (2016).
\newblock \enquote{Subsampling Versus Bootstrapping in Resampling-Based Model
  Selection for Multivariable Regression.}
\newblock \emph{Biometrics}, \textbf{72}(1), 272--280.

\bibitem[{Dembczy\'nski \emph{et~al.}(2008)Dembczy\'nski, Kot{\l}owski, and
  S{\l}owi\'nski}]{DembyKotl08}
Dembczy\'nski K, Kot{\l}owski W, S{\l}owi\'nski R (2008).
\newblock \enquote{Solving Regression by Learning an Ensemble of Decision
  Rules.}
\newblock In \emph{International Conference on Artificial Intelligence and Soft
  Computing, 2008}, pp. 533--544. Springer-Verlag, Heidelberg, Germany.

\bibitem[{Dembczy{\'n}ski \emph{et~al.}(2010)Dembczy{\'n}ski, Kot{\l}owski, and
  S{\l}owi{\'n}ski}]{DembyKotl10}
Dembczy{\'n}ski K, Kot{\l}owski W, S{\l}owi{\'n}ski R (2010).
\newblock \enquote{ENDER: A Statistical Framework for Boosting Decision Rules.}
\newblock \emph{Data Mining and Knowledge Discovery}, \textbf{21}(1), 52--90.
\newblock \doi{10.1007/s10618-010-0177-7}.

\bibitem[{Deng(2014)}]{Deng14}
Deng H (2014).
\newblock \emph{\pkg{inTrees}: Interpret Tree Ensembles}.
\newblock \proglang{R} package version 1.1,
  \urlprefix\url{https://CRAN.R-project.org/package=inTrees}.

\bibitem[{Dietterich(2000)}]{Diet00}
Dietterich T (2000).
\newblock \enquote{Ensemble Methods in Machine Learning.}
\newblock In J~Kittler, F~Roli (eds.), \emph{Multiple Classifier Systems}, pp.
  1--15. Springer-Verlag, Berlin, Germany.

\bibitem[{Dua and Graff(2019)}]{DuayGraf19}
Dua D, Graff C (2019).
\newblock \enquote{{UCI} Machine Learning Repository.}
\newblock \urlprefix\url{http://archive.ics.uci.edu/ml}.

\bibitem[{Efron and Tibshirani(1994)}]{EfroyTibs94}
Efron B, Tibshirani RJ (1994).
\newblock \emph{An Introduction to the Bootstrap}.
\newblock CRC press, New York, NY.

\bibitem[{Fern{\'a}ndez-Delgado \emph{et~al.}(2014)Fern{\'a}ndez-Delgado,
  Cernadas, Barro, and Amorim}]{FernyCern14}
Fern{\'a}ndez-Delgado M, Cernadas E, Barro S, Amorim D (2014).
\newblock \enquote{Do We Need Hundreds of Classifiers to Solve Real World
  Classification Problems?}
\newblock \emph{Journal of Machine Learning Research}, \textbf{15}(1),
  3133--3181.

\bibitem[{Fokkema and Christoffersen(2020)}]{FokkyChri19}
Fokkema M, Christoffersen B (2020).
\newblock \emph{\pkg{pre}: Prediction Rule Ensembles}.
\newblock \proglang{R} package version 1.0.0,
  \urlprefix\url{https://CRAN.R-project.org/package=pre}.

\bibitem[{Fokkema \emph{et~al.}(2015)Fokkema, Smits, Kelderman, and
  Penninx}]{FokkySmit15a}
Fokkema M, Smits N, Kelderman H, Penninx BW (2015).
\newblock \enquote{Connecting Clinical and Actuarial Prediction With Rule-Based
  Methods.}
\newblock \emph{Psychological Assessment}, \textbf{27}(2), 636--644.
\newblock \doi{10.1037/pas0000072}.

\bibitem[{Frank and Witten(1998)}]{FranyWitt98}
Frank E, Witten IH (1998).
\newblock \enquote{Generating Accurate Rule Sets Without Global Optimization.}
\newblock In \emph{Proceedings of the Fifteenth International Conference on
  Machine Learning}, pp. 144--151. Morgan Kaufmann Publishers, San Mateo, CA.

\bibitem[{Friedman(2001)}]{Frie01}
Friedman J (2001).
\newblock \enquote{Greedy Function Approximation: A Gradient Boosting Machine.}
\newblock \emph{Annals of Statistics}, \textbf{29}(5), 1189--1232.
\newblock \doi{10.1214/aos/1013203451}.

\bibitem[{Friedman \emph{et~al.}(2010)Friedman, Hastie, and
  Tibshirani}]{FrieyHast10}
Friedman J, Hastie T, Tibshirani R (2010).
\newblock \enquote{Regularization Paths for Generalized Linear Models via
  Coordinate Descent.}
\newblock \emph{Journal of Statistical Software}, \textbf{33}(1), 1--22.
\newblock \doi{10.18637/jss.v033.i01}.

\bibitem[{Friedman and Popescu(2003)}]{FrieyPope03}
Friedman J, Popescu B (2003).
\newblock \enquote{Importance Sampled Learning Ensembles.}
\newblock \emph{Technical report}, Stanford University.
\newblock \urlprefix\url{http://www-stat.stanford.edu/~jhf/ftp/isle.pdf}.

\bibitem[{Friedman and Popescu(2008)}]{FrieyPope08}
Friedman J, Popescu B (2008).
\newblock \enquote{Predictive Learning via Rule Ensembles.}
\newblock \emph{The Annals of Applied Statistics}, \textbf{2}(3), 916--954.
\newblock \doi{10.1214/07-aoas148}.

\bibitem[{Friedman and Popescu(2012)}]{FrieyPope12}
Friedman J, Popescu B (2012).
\newblock \enquote{RuleFit with \proglang{R} [version 3].}
\newblock \urlprefix\url{http://www-stat.stanford.edu/~jhf/R-RuleFit.html}.

\bibitem[{F{\"u}rnkranz(1999)}]{Furn99}
F{\"u}rnkranz J (1999).
\newblock \enquote{Separate-And-Conquer Rule Learning.}
\newblock \emph{Artificial Intelligence Review}, \textbf{13}(1), 3--54.
\newblock \doi{10.1023/a:1006524209794}.

\bibitem[{Hall \emph{et~al.}(2009)Hall, Frank, Holmes, Pfahringer, Reutemann,
  and Witten}]{HallyFran09}
Hall M, Frank E, Holmes G, Pfahringer B, Reutemann P, Witten IH (2009).
\newblock \enquote{The WEKA Data Mining Software: An Update.}
\newblock \emph{ACM SIGKDD Explorations Newsletter}, \textbf{11}(1), 10--18.

\bibitem[{Holmes \emph{et~al.}(1999)Holmes, Hall, and Frank}]{HolmyHall99}
Holmes G, Hall M, Frank E (1999).
\newblock \enquote{Generating Rule Sets from Model Trees.}
\newblock In N~Foo (ed.), \emph{Proceedings of the Twelfth Australian Joint
  Conference on Artificial Intelligence}, pp. 1--12. Springer-Verlag,
  Heidelberg, Germany.

\bibitem[{Hothorn \emph{et~al.}(2006)Hothorn, Hornik, and
  Zeileis}]{HothyHorn06}
Hothorn T, Hornik K, Zeileis A (2006).
\newblock \enquote{Unbiased Recursive Partitioning: A Conditional Inference
  Framework.}
\newblock \emph{Journal of Computational and Graphical Statistics},
  \textbf{15}(3), 651--674.
\newblock \doi{10.1198/106186006x133933}.

\bibitem[{Hothorn \emph{et~al.}(2005)Hothorn, Leisch, Zeileis, and
  Hornik}]{HothyLeis05}
Hothorn T, Leisch F, Zeileis A, Hornik K (2005).
\newblock \enquote{The Design and Analysis of Benchmark Experiments.}
\newblock \emph{Journal of Computational and Graphical Statistics},
  \textbf{14}(3), 675--699.
\newblock \doi{10.1198/106186005x59630}.

\bibitem[{Hothorn and Zeileis(2015)}]{HothyZeil15}
Hothorn T, Zeileis A (2015).
\newblock \enquote{\pkg{partykit}: A Modular Toolkit for Recursive Partytioning
  in \proglang{R}.}
\newblock \emph{Journal of Machine Learning Research}, \textbf{16}, 3905--3909.

\bibitem[{Joly \emph{et~al.}(2012)Joly, Schnitzler, Geurts, and
  Wehenkel}]{JolyySchn12}
Joly A, Schnitzler F, Geurts P, Wehenkel L (2012).
\newblock \enquote{L1-Based Compression of Random Forest Models.}
\newblock In \emph{20th European Symposium on Artificial Neural Networks,
  Computational Intelligence and Machine Learning}. Bruges, Belgium.

\bibitem[{Kuhn and Johnson(2013)}]{KuhnyJohn13}
Kuhn M, Johnson K (2013).
\newblock \emph{Applied Predictive Modeling}.
\newblock Springer-Verlag, New York, NY.

\bibitem[{Kuhn \emph{et~al.}(2012)Kuhn, Weston, Coulter, Culp, and
  Quinlan}]{Kuhn15}
Kuhn M, Weston S, Coulter N, Culp M, Quinlan R (2012).
\newblock \emph{\pkg{C50}: C5.0 Decision Trees and Rule-Based Models}.
\newblock \proglang{R} package version 0.1.0,
  \urlprefix\url{https://CRAN.R-project.org/package=C50}.

\bibitem[{Leisch and Dimitriadou(2012)}]{LeisyDimi12}
Leisch F, Dimitriadou E (2012).
\newblock \emph{\pkg{mlbench}: Machine Learning Benchmark Problems}.
\newblock \proglang{R} package version 2.1-1,
  \urlprefix\url{https://CRAN.R-project.org/package=mlbench}.

\bibitem[{Liaw and Wiener(2002)}]{LiawyWien02}
Liaw A, Wiener M (2002).
\newblock \enquote{Classification and Regression by \pkg{randomForest}.}
\newblock \emph{\proglang{R} News}, \textbf{2}(3), 18--22.
\newblock \urlprefix\url{http://CRAN.R-project.org/doc/Rnews/}.

\bibitem[{Lin and Jeon(2006)}]{LinyJeon06}
Lin Y, Jeon Y (2006).
\newblock \enquote{Random Forests and Adaptive Nearest Neighbors.}
\newblock \emph{Journal of the American Statistical Association},
  \textbf{101}(474), 578--590.
\newblock \doi{10.1198/016214505000001230}.

\bibitem[{Meinshausen(2010)}]{Mein10}
Meinshausen N (2010).
\newblock \enquote{Node Harvest.}
\newblock \emph{The Annals of Applied Statistics}, \textbf{4}(4), 2049--2072.
\newblock \doi{10.1214/10-aoas367}.

\bibitem[{Nalenz and Villani(2017)}]{NaleyVill17pkg}
Nalenz M, Villani M (2017).
\newblock \emph{\pkg{horserule}: Flexible Non-Linear Regression with the
  HorseRule Algorithm}.
\newblock \proglang{R} package version 0.1.0,
  \urlprefix\url{https://CRAN.R-project.org/package=horserule}.

\bibitem[{Quinlan(1992)}]{Quin92}
Quinlan JR (1992).
\newblock \enquote{Learning With Continuous Classes.}
\newblock In \emph{5th Australian Joint Conference on Artificial Intelligence},
  volume~92, pp. 343--348. World Scientific, Singapore.

\bibitem[{Quinlan(1993)}]{Quin93}
Quinlan JR (1993).
\newblock \emph{{\pkg{C4.5}: Programs for Machine Learning}}.
\newblock Morgan Kaurfmann, San Mateo, CA.

\bibitem[{{\proglang{R} Core Team}(2019)}]{R19}
{\proglang{R} Core Team} (2019).
\newblock \emph{\proglang{R}: A Language and Environment for Statistical
  Computing}.
\newblock \proglang{R} Foundation for Statistical Computing, Vienna, Austria.
\newblock \urlprefix\url{https://www.R-project.org/}.

\bibitem[{RuleQuest(2017)}]{Rule17}
RuleQuest (2017).
\newblock \emph{"Is See5/C5.0 better than C4.5?"}.
\newblock RuleQuest.
\newblock \urlprefix\url{https://rulequest.com/see5-comparison.html}.

\bibitem[{Schauerhuber \emph{et~al.}(2007)Schauerhuber, Zeileis, Meyer, and
  Hornik}]{SchayZeil07}
Schauerhuber M, Zeileis A, Meyer D, Hornik K (2007).
\newblock \enquote{Benchmarking Open-Source Tree Learners in
  \proglang{R}/\pkg{RWeka}.}
\newblock In C~Preisach, H~Burkhardt, L~Schmidt-Thieme, R~Decker (eds.),
  \emph{Data Analysis, Machine Learning and Applications}, pp. 389--396.
  Springer-Verlag, Heidelberg, Germany.

\bibitem[{Shimokawa \emph{et~al.}(2014)Shimokawa, Li, Yan, Kitamura, and
  Goto}]{ShimyLi14}
Shimokawa T, Li L, Yan K, Kitamura S, Goto M (2014).
\newblock \enquote{Modified Rule Ensemble Method for Binary Data and its
  Applications.}
\newblock \emph{Behaviormetrika}, \textbf{41}(2), 225--244.
\newblock \doi{10.2333/bhmk.41.225}.

\bibitem[{Strobl \emph{et~al.}(2009)Strobl, Malley, and Tutz}]{StroyMall09}
Strobl C, Malley J, Tutz G (2009).
\newblock \enquote{An Introduction to Recursive Partitioning: Rationale,
  Application, and Characteristics of Classification and Regression Trees,
  Bagging, and Random Forests.}
\newblock \emph{Psychological Methods}, \textbf{14}(4), 323.
\newblock \doi{10.1037/a0016973}.

\bibitem[{Therneau \emph{et~al.}(2019)Therneau, Atkinson, and
  Ripley}]{TheryAtki19}
Therneau T, Atkinson B, Ripley B (2019).
\newblock \emph{\pkg{rpart}: Recursive Partitioning and Regression Trees}.
\newblock R package version 4.1-15,
  \urlprefix\url{https://CRAN.R-project.org/package=rpart}.

\bibitem[{Weiss and Indurkhya(2000)}]{WeisyIndu00}
Weiss SM, Indurkhya N (2000).
\newblock \enquote{Lightweight Rule Induction.}
\newblock In \emph{Proceedings of the Seventeenth International Conference on
  Machine Learning}, pp. 1135--1142. Morgan Kaufmann Publishers Inc., San
  Mateo, CA.

\bibitem[{Yang \emph{et~al.}(2008)Yang, Zhang, Chen, Chen, Li, and
  Lu}]{YangyZhan08}
Yang W, Zhang S, Chen Y, Chen Y, Li W, Lu H (2008).
\newblock \enquote{{Mining Diagnostic Rules of Breast Tumor on Ultrasound Image
  Using Cost-Sensitive RuleFit Method}.}
\newblock In \emph{International Conference on Intelligent System and Knowledge
  Engineering (ISKE 2008)}, pp. 354--359. IEEE, Xiamen, China.

\bibitem[{Zeileis \emph{et~al.}(2008)Zeileis, Hothorn, and
  Hornik}]{ZeilyHoth08}
Zeileis A, Hothorn T, Hornik K (2008).
\newblock \enquote{Model-Based Recursive Partitioning.}
\newblock \emph{Journal of Computational and Graphical Statistics},
  \textbf{17}(2), 492--514.
\newblock \doi{10.1198/106186008x319331}.

\end{thebibliography}

\end{document}